\documentclass[usenatbib,useAMS,usegraphicx]{mn2e}
\usepackage{rotating}
\usepackage{float}
\usepackage{subfigure}
\usepackage{graphicx}
\usepackage{amsmath}
\usepackage{color}

\newcommand{\vect}{\bmath}
\newcommand{\ufr}{$0.0076_{-0.0052}^{+0.0208}$}
\newcommand{\ual}{$< 2.93$}
\newcommand{\gfr}{$0.0064^{+0.0080}_{-0.0042}$}
\newcommand{\gal}{$1.90^{+0.098}_{-0.098}$}
\newcommand{\Evu}{-5.98}
\newcommand{\Evg}{-6.13}

\title[CDM substructure mass function  at $z=0.2$]{Inference of the cold dark matter substructure mass function  at $z=0.2$ using strong gravitational lenses\\} 
\author[Vegetti et al.]{S.
  Vegetti$^{1,2}$, L.V.E. Koopmans$^{3}$, M.W. Auger$^4$, T. Treu$^5$, A.S. Bolton$^6$\\
  $^1$ Max Planck Institute for Astrophysics, Karl-Schwarzschild-Strasse 1, 85740 Garching, Germany\\
  $^2$ Kavli Institute for Astrophysics and Space Research, Massachusetts Institute of Technology, Cambridge, MA 02139, USA\\
  $^3$ Kapteyn Astronomical Institute, University of Groningen, PO Box 800, 9700 AV Groningen, the Netherlands\\
  $^4$ Institute of Astronomy, Madingley Road, Cambridge CB3 0HA, UK\\
  $^5$ Department of Physics, University of California, Santa Barbara, CA 93106, USA\\
  $^6$ Department of Physics and Astronomy, University of Utah, Salt Lake City, UT 84112, USA}
\begin{document}

\maketitle

\begin{abstract}
We present the results of a search for galaxy substructures in a sample of 11 gravitational lens galaxies from the Sloan Lens ACS Survey. We find no significant detection of mass clumps, except for a luminous satellite in the system SDSS\,J0956+5110. 
We use these non-detections, in combination with a previous detection in the system SDSS\,J0946+1006, to derive constraints on the substructure mass function in massive early-type host galaxies with an average redshift $\langle z_{\rm lens} \rangle\sim0.2$ and an average velocity dispersion $\langle \sigma_{\rm{eff}}\rangle\sim 270$ km\,s$^{-1}$. 
We perform a Bayesian inference on the substructure mass function, within a median region of about 32 kpc$^2$ around the Einstein radius ($\langle R_{\rm{ein}}\rangle\sim 4.2$ kpc). We infer a mean projected substructure mass fraction $f=~$\ufr ~at the 68 percent confidence level and a substructure mass function slope $\alpha ~$\ual ~at the 95 percent confidence level for a uniform prior probability density on $\alpha$. For a Gaussian prior based on cold dark matter (CDM) simulations, we infer $f =~$\gfr  ~and a slope of $\alpha = ~$\gal ~at the 68 percent confidence level. Since only one substructure was detected in the full sample, we have little information on the mass function slope, which is therefore poorly constrained (i.e. the Bayes factor shows no positive preference for any of the two models). The inferred fraction is consistent with the expectations from CDM simulations and with inference from flux ratio anomalies at the 68 percent confidence level. 
\end{abstract}


\section{Introduction}
One of the most robust predictions of the cold dark matter (CDM) theory is that dark matter haloes of all scales should be populated by thousands of low-mass substructures \citep{Kauffmann93,Klypin99,Diemand08,Springel08}.
While an equivalent number of satellites has not yet been detected in observations of the Local Group galaxies \citep[e.g.][and references therein]{Kravtsov10}, reconciling the observations with the theoretical expectations could be possible by invoking the existence of a large population of low surface brightness or even dark satellites. Alternatively, solving the problem would require modifying the nature of dark matter \citep[e.g.][]{Nierenberg13}.

Strong gravitational lensing, being sensitive to any form of matter, provides a unique opportunity to address this issue.
It was initially suggested by \citet{Mao98} that flux-ratio anomalies of multiply imaged lensed quasars can be interpreted as the signature of mass substructure in the lens galaxies at scales smaller than the lensed image separation. Indeed, \citet{Metcalf01}, \citet{Chiba02}, \citet{Dalal02}, \citet{Metcalf02}, \citet{Keeton03} and \citet{Nierenberg14} showed that mass substructure could be responsible for a violation of the magnification relations. 

While early comparisons with $N$-body numerical simulations have indicated an agreement between the predicted and the gravitational lensing inferred amount of substructure \citep[e.g.][]{Bradac02}, more recently \citet{Xu09} have shown that the probability of reproducing the observed rate of anomalies with the subhalo population of the Aquarius simulation is of the order of $10^{-3}$. The inclusion of line-of-sight structures from the Millennium II simulations only increases the chance of flux-ratio anomalies to about 10 - 30 percent \citep{Xu12}.
At the same time however, gravitational lens galaxies are more commonly massive early-type galaxies and are hosted therefore by dark matter haloes that are more massive and potentially different (e.g. with a higher ellipticity) than those of the Aquarius simulation, which are primarily Milky-Way-type haloes. In particular, \citet{Metcalf12} using more realistic simulations of lens galaxies have found consistent results between the expectations and the observations. A similar conclusion was more recently obtained by \citet{Xu13}.

\citet{Schechter93}, \citet{McKean07} and \citet{More09} have found that in some cases, the flux-ratio anomaly is related to the presence of a single massive luminous satellite rather than a large population of small mass CDM substructures. It has also been suggested that evidence for mass substructure in lens galaxies can be obtained from astrometric measurements alone \citep{Chen07} or in combination with flux-ratio anomalies \citep{Fadely12} and time-delay measurements \citep{Keeton09}.

\citet{Koopmans05} and \citet{Vegetti09a} have developed an alternative technique of direct gravitational imaging of mass substructure based on perturbations to the surface brightness distribution of lensed arcs and Einstein rings. 
While the gravitational imaging technique is complementary to the above methods, it has the major advantage of being able to directly measure the substructure mass and position, and therefore is less degenerate in the mass models. This feature of the gravitational imaging technique makes it particularly powerful in constraining the substructure mass function \citep{Vegetti09b}.

The gravitational imaging technique has previously been applied to high resolution {\it Hubble Space Telescope} ({\it HST}; \citealt{Vegetti10}) and Keck adaptive optics imaging \citep{Vegetti12}. This method has already led to the detection of two substructures at redshifts of 0.222 and 0.881, with respective masses of $(3.51\pm 0.15)\times10 ^9M_\odot$ and $(1.9 \pm 0.1)\times 10^8 M_\odot$. The substructure mass here quoted is the total 3D mass measured under the assumption of a spherical pseudo-Jaffe mass profile (Section \ref{clumpy_model}). As shown in \citet{Vegetti12} the mass so measured is consistent with the projected mass derived in a model-independent fashion from the pixelized density corrections (Section \ref{pot_corr}). This indicates that the measured lensing mass is not significantly affected by our choice of mass density profile \citep[see also][]{Vegetti14}. Under the assumption that these detections are associated with the host lens galaxy, they imply a substructure projected (within $\sim$ 5 kpc) mass fraction of $\langle f \rangle =0.033^{+0.036}_{-0.018}$ and a slope of the substructure mass function of $\alpha=1.1^{+0.6}_{-0.4}$ (where $dN/dm\propto m^{-\alpha}$). Within the large errors, these results are consistent with the theoretical expectations at the 95 percent confidence level \citep[CL;][]{Springel08, Diemand08, Xu13}. The mass fraction is inferred by assuming a mass function normalized between $4\times10^6 M_\odot$ and $4\times10^9 M_\odot$.

In this paper, we present a statistical analysis of a sample of 11 gravitational lens galaxies with the aim of deriving tighter constraints on the substructure mass function.
The layout of the paper is as follows. In Section 2 we provide a short description of the analysed data. In Section 3 we provide an overview of the full analysis from the lens modelling to the derivation of the substructure mass function. In Section 4, we describe the gravitational imaging technique, while in Section 5 we present the main lens modelling results. In Section 6, we derive statistical constraints on the substructure mass function and we finally conclude in Section 7.

Throughout the paper we assume the following cosmology, $H_{\rm{0}} = 73\,\rm{km\,s^{-1}}~\rm{Mpc}^{-1}$, $\Omega_{\rm m}=0.25$ and $\Omega_\Lambda=0.75$. 

\section{Data}
The gravitational lenses analysed in this paper are all taken from the Sloan Lens ACS Survey \citep[SLACS, e.g.][]{Bolton08a}. These were selected from the main \citep{Strauss02} and luminous red galaxy (LRG; \citealt{Eisenstein01}) samples of the Sloan Digital Sky Survey (SDSS; \citealt{York00}). The SLACS gravitational lenses are generally massive early-type galaxies, typical of the parent samples from which they are selected \citep[e.g.][]{Bolton06}.

The data used in the current analysis were obtained using the wide-field channel (WFC) of the Advanced Camera for Surveys (ACS) aboard the \textsl{HST}. A single orbit was used for each unique filter, split into four dithered exposures.  The data were rectified and co-added on to a uniform image grid using custom software written in the IDL language. The surface brightness profiles of the lensing foreground galaxies were modelled and subtracted using the radial B-spline formalism described by \citet{Bolton06,Bolton08a}. Initial models for the lens galaxy surface brightness distribution were optimized non-linearly to determine the centroid, axis ratio and position angle of the lens galaxies. These parameters were then used to define an elliptical coordinate system for the final linear model fitting, including sufficient multipole orders to provide model-subtracted images free of appreciable systematic residuals. Masks for the lensed image features were generated manually and applied during the B-spline model fitting. In some cases, these masks were updated based on initial B-spline subtracted residual images, and the fitting procedures were repeated. \citet{Vegetti10}  have explicitly tested the effect of the galaxy subtraction procedure on the significance of substructure detections using a S\'{e}rsic profile rather than a B-spline surface brightness profile and found no significant influence. Moreover,  the B-spline subtraction was done independently for each filter and we find in this paper that the sensitivity function to substructures (see Section \ref{sensitivity}) is consistent across filters. A similar test with an equivalent output was done by \citet{Vegetti12}. We are confident, therefore, that the galaxy subtraction procedure is not a major source of systematic error on our inference of the substructure mass function. This can be understood because the effect of a substructure on the brightness distribution of a lensed image occurs on spatial scales over which the galaxy surface brightness model varies very little (i.e. there is very little mixing of spatial scales and early-type galaxies show very little structure on the scales where substructure causes anomalies).
 
 In order to increase our sensitivity to lower mass substructures, the subsample of lenses investigated in this paper was selected purely on the basis of the lensed images signal-to-noise ratio ($S/N$) in the $I$ and $V$ band, whenever ACS images are available. Specifically, all the images in the full 11 lens sample have at least 200 pixels with a $S/N\ge 2$ and at least 50 pixels with $S/N\ge 3$. This ensures that there is a region on the lensed images large enough for substructures to be detected when present with a mass above the detection threshold. This selection does not bias in favour or against substructure in the lens, since substructure in the probed mass-range has a negligible effect on the source magnification. Since no substructure detection was obtained for the sample here considered, the lack of {\it V}-band data for some of the systems does not represent a major issue for the interpretation of the lens modelling.
When constraining the substructure mass function, we also consider our previous detection in the gravitational lens system SDSS\,J0946+1006 that was similarly analysed by \citet{Vegetti10}. In Table \ref{tbl:lenses} we list all of the lens systems in our sample with the lens and source redshifts, the filter used for the imaging, and other relevant properties. The gravitational lens system B1938+666, previously analysed by \citet{Vegetti12}, is not included in the current sample as this system is not part of the SLACS survey but of the Cosmic Lens All-Sky Survey (CLASS), and follows therefore a different selection function (source based rather than lensed based). Moreover, since the amount of substructure in lens galaxies can significantly depend on the properties of the host galaxy \citep[e.g.][]{Xu13}, by limiting our sample to the SLACS lenses, we can avoid possible issues related to trends of the substructure mass fraction with the host redshift and mass.

\begin{table*}
\caption{The list of the gravitational lens systems considered in this paper, with the lens and source redshift, the lens effective radius, the velocity dispersion within half of the effective radius, the stellar mass \citep[from][]{Auger09} and the filter used for their imaging, and for the lens modelling ($I =$ F{\rm{814W}} and $V =$ F{\rm{555W}}).}
\begin{tabular}{ccccccc}
\hline
Name&$z_{\rm{lens}}$&$z_{\rm{source}}$&$R_{\rm{eff}}$&$\sigma_{\rm{eff/2}}$&$\log[M_{*}^{\rm{Salp}}/M_\odot]$&Filter\\
\small{(SDSS)}&&&\small{[kpc]}&\small{[km\,s$^{-1}$]}\\
\hline

$\rm{J0252+0039}$&0.280&0.982&5.68&170$\pm$12&11.46$\pm$0.13&$I$\\

$\rm{J0737+3216}$&0.322&0.581&14.10&338$\pm$16&11.96$\pm$0.07&$I$ \& $V$\\

$\rm{J0946+1006}$&0.222&0.609&9.08&256$\pm$21&11.59$\pm$0.12&$I$\\

$\rm{J0956+5100}$&0.240&0.470&8.58&338$\pm$15&11.81$\pm$0.08&$I$ \& $V$\\

$\rm{J0959+4416}$&0.237&0.531&7.27&248$\pm$19&11.72$\pm$0.12&$I$\\

$\rm{J1023+4230}$&0.191&0.696&5.97&247$\pm$15&11.57$\pm$0.12&$I$\\

$\rm{J1205+4910}$&0.215&0.481&9.04&282$\pm$13&11.72$\pm$0.06&$I$ \& $V$\\

$\rm{J1430+4105}$&0.285&0.575&10.41&325$\pm$32&11.93$\pm$0.11&$I$\\

$\rm{J1627-0053}$&0.208&0.524&6.87&295$\pm$14&11.70$\pm$0.09&$I$ \& $V$\\

$\rm{J2238-0754}$&0.137&0.713&5.78&200$\pm$11&11.45$\pm$0.06&$I$ \& $V$\\

$\rm{J2300+0022}$&0.228&0.463&6.88&284$\pm$17&11.65$\pm$0.07&$I$ \& $V$\\

\hline
\end{tabular}
\label{tbl:lenses}
\end{table*}

\section{Analysis overview}
We summarize here all of the steps involved in our analysis and we refer to the following sections and to \citet{Koopmans05} and \citet{Vegetti09a, Vegetti09b} for details on the lens modelling technique and the statistical inference on the substructure mass function:

\begin{enumerate}
\item  all of the lens galaxies in the sample are initially modelled under the assumption of a smooth mass distribution with an elliptical power-law density profile and with the inclusion of external shear (Section \ref{smooth_model});\\
\item for each lens, the smooth model is improved with linear and localized potential corrections, which are examined for the presence of mass substructures (Section \ref{pot_corr});\\
\item all of the lens systems that have shown evidence for non-negligible (at the 10$\sigma$ level) potential corrections are remodelled with the same smooth mass model family of step (i), plus one or more mass substructures with a given analytical mass density profile (Section \ref{clumpy_model});\\
\item for each lens we calculate the 10$\sigma$ mass-detection threshold at each pixel in the lens plane and derive the mass sensitivity as a function of image-plane position (Section \ref{sensitivity}, Figure \ref{fig:sens_func});\\
\item we determine the joint likelihood for the set of detections and non-detections and  then transform it, in a Bayesian way, into posterior probability density functions (PDFs) for the CDM substructure mass fraction and mass function slope (Section \ref{sec:posterior_mf}, Figure \ref{fig:mass_function}).
\end{enumerate}

\section{Lens modelling technique}
The gravitational lens modelling is carried out in a three step procedure using the Bayesian grid-based technique of \citet{Vegetti09a}. Below, we describe each step of this technique in more detail. 

\subsection{Smooth models}
\label{smooth_model}
We assume the projected mass density profile of the lens to be well described by an elliptical power-law distribution plus an external shear, with a dimensionless surface density and Einstein radius defined, respectively, as 
\begin{equation}
\kappa(x,y)= \frac{\kappa_0\left(2-\frac{\gamma}{2}\right)q^{\gamma-3/2}}{2(q^2(x^2+r_{\rm c}^2)+y^2)^{(\gamma -1) /2}}\,
\label{equ:kappa_smooth}
\end{equation}
and 
\begin{equation}
R_{\rm{ein}} =\left(\frac{\kappa_0 (2-\gamma/2) q^{(\gamma-2)/2}}{ 3 - \gamma}\right)^{1/(\gamma-1)}\,.
\end{equation}

\noindent For the isothermal case  ($\gamma=2$) this is the same expression introduced by \citet{Kormann94}. The normalization $\kappa_0\left(2-\frac{\gamma}{2}\right)q^{\gamma-3/2}$ is chosen such that the mass within the tangential critical curve is independent of the axis ratios $q$ for a fixed velocity dispersion and  to reduce the degeneracy between $\kappa_0$ and the mass density slope $\gamma$. In the limit of a singular, spherical ($q=1$) and isothermal model,  $\kappa_0$ is the Einstein radius (in the units of $x$ and $y$) of the model and is related to the 1-d velocity dispersion of the singular isothermal sphere as defined in \citet{Binney87}. 

\noindent The free parameters of this mass model are the mass density normalization $\kappa_0$, its position angle $\theta$, the axis ratio $q$, the coordinate centre $x$ and $y$, the logarithmic slope $\gamma$, the external shear strength $\Gamma$ and its position angle $\Gamma_\theta$. The mass distribution is assumed to have a negligible core radius (i.e. $r_{\rm c}\equiv 0$ arcsec). \\

\subsubsection{Model optimization and inference}
In this section, we shortly outline our Bayesian model inference.
Given the surface brightness distribution of the lensed images $\bmath d$ with noise $\bmath n$ and an unknown source surface brightness distribution $\bmath s$, the most probable a posteriori set of mass model parameters $\bmath \eta = \{\kappa_0,\theta,q,x,y,\gamma,\Gamma,\Gamma_{\theta}\}$ and the source regularization parameter $\lambda_{\rm{s}}$ are obtained by solving the set of linear equations:
  \begin{equation}
    \mathbf{M_c}\left(\bmath{\eta},\bmath\psi\right)\,\bmath s = \bmath d +
    \bmath n\,.
    \label{equ: src_pot_linear_blurred}
  \end{equation} 
Here, $\mathbf{M_c}\left(\bmath{\eta},\bmath\psi\right)$ is the response matrix and accounts for the linear lensing operator and for the point spread function (PSF) blurring. The lensing operator is a non-linear function of the parameters $\bmath \eta$ via the lensing potential $\nabla^{2} \psi(\bmath x)=2\kappa(\bmath x)$. In practice the above equations are solved by maximizing this posterior probability density distribution: 
\begin{equation}\label{equ:posterior_2}
    P(\lambda_{\rm{s}},\bmath \eta\,|\,\bmath d,\mathbf{M_c},\mathbf
      R_{\rm{s}})=\frac{P(\bmath d\,|\,\lambda_{\rm{s}},\bmath \eta,\mathbf{M_c},\mathbf
      R_{\rm{s}})P(\lambda_{\rm{s}},\bmath \eta)}{P(\bmath d\,|\,\mathbf{M_c},\mathbf
      R_{\rm{s}})}\,.
\end{equation}
In the above expression, $\mathbf {R_{\rm{s}}}$ encodes the source regularization form (e.g. variance, gradient or curvature), the strength of which is set by the source regularization level $\lambda_{\rm{s}}$ \citep[see][for details on these operators]{Koopmans05, Vegetti09a}. The optimization is performed in three separate steps: first, $\lambda_{\rm{s}}$ is kept fixed at a relatively large value, such that the source model remains relatively smooth, and $P(\bmath \eta\,|\,\lambda_{\rm{s}},\bmath d,\mathbf{M_c},\mathbf R_{\rm{s}})$ is maximized by varying $\bmath \eta$; secondly, the lens parameters are kept fixed at their most probable values found during the first step, while $P(\lambda_{\rm{s}}\,|\,\bmath \eta, \bmath d,\mathbf{M_c},\mathbf R_{\rm{s}})$ is optimized for the source regularization level $\lambda_{\rm{s}}$; finally,  $P(\bmath \eta\,|\,\lambda_{\rm{s}},\bmath d,\mathbf{M_c},\mathbf R_{\rm{s}})$ is maximized again for the lens parameters with a source regularization level fixed at the most probable value determined from the second step. In general, no further iterations are needed. 

At every step of the non-linear mass model optimization, the corresponding most probable source surface brightness distribution $\bmath s$ is obtained, under the assumption of Gaussian noise\footnote{The noise in each pixel includes background noise from the sky and the electronics as well as shot noise. The DRIZZLE scale and PIXFRAC were chosen so that the pixel-to-pixel correlations are very small.}, by maximizing the following probability density  through the direct inversion of the linear equation given in equation (\ref{equ: src_pot_linear_blurred}):
\begin{equation}
    P\left(\bmath s\,|\,\bmath d,\lambda_{\rm{s}}\bmath \eta,\mathbf
    {M_c},\mathbf R_{\rm{s}}\right)=\frac{P(\bmath d \,|\,\bmath s,\bmath \eta,
    \mathbf{M_c})\, P(\bmath s\,|\,\lambda_{\rm{s}},\mathbf R_{\rm{s}})}{P(\bmath 
    d \,|\,\lambda_{\rm{s}},\bmath \eta,\mathbf{M_c},\mathbf R_{\rm{s}})}\,.
   \label{eq:evidence} 
 \end{equation}
\noindent Specifically, the source surface brightness distribution is constructed on a Delaunay tessellation with a resolution that is adaptive with the lensing magnification and that is built by casting one pixel from each contiguous block of $n_{\rm{src}}\times n_{\rm{src}}$ pixels from the image plane to the source plane via the lens equation. The lower that $n_{\rm{src}}$ is, the higher the source grid resolution.
The optimal (i.e. the one that maximizes the evidence of the data given the model) number of pixels as well as the best form of source regularization can be different from system to system (often depending on the smoothness of the lensed images). For example, thin  rings with a large dynamic range in their surface brightness distribution often require a higher source resolution (lower $n_{\rm{src}}$) and an adaptive source regularization (i.e. a regularization that changes from pixel to pixel according to the specific S/N) in order to fit all of the features in the image structure. Similarly, \citet{Suyu06} showed that different source surface brightness distributions may require different forms of regularization. The best combination of $n_{\rm{src}}$ and $\mathbf R_{\rm{s}}$ for every lens is determined by re-running the smooth-lens three-step optimizing procedure, described above, for an increasing number of grid pixels and for different forms of regularization, either curvature or gradient, either adaptive or non-adaptive. While our choice of $n_{\rm{src}}$ and $\mathbf R_{\rm{s}}$ does not significantly affect the recovered mass-model parameters, ${\bmath \eta}$, it can have a substantial effect on the level with which our best model matches the original data. Different choices for these parameters are compared in terms of the Bayesian evidence, which allows us to objectively determine their optimal values. This overall procedure effectively removes any choice of the astronomer in modelling these lenses that (unconsciously) influence the results.

\subsection{Lens-potential corrections}
\label{pot_corr}
The gravitational imaging technique allows the detection of mass substructure in lens galaxies as linear localized and pixelized corrections to the overall smooth analytic lens potential, as described in Section \ref{smooth_model} \citep{Koopmans05, Vegetti09a}. Through a Gauss-Newton optimization scheme, the maximum a posteriori (MAP) corrected lens potential $\psi(\bmath x)+\delta\psi(\bmath x)$, is computed  for several levels of potential correction regularization, along with the corresponding MAP source model, by solving the following set of equations for $\bmath r = (\bmath s ,\delta\bmath\psi)$:
\begin{equation}
    P\left(\bmath r\,|\,\bmath d,\bmath\lambda,\bmath \eta,\mathbf
    {M_c},\mathbf R\right)=\frac{P(\bmath d \,|\,\bmath r,\bmath \eta,
    \mathbf{M_c})\, P(\bmath r\,|\,\bmath\lambda,\mathbf R)}{P(\bmath
    d \,|\,\bmath\lambda,\bmath \eta,\mathbf{M_c},\mathbf R)}\,.
  \end{equation}
\noindent Here, $\mathbf{M_c}$ is a function of $\psi(\bmath x)$ (which is updated by $\delta\psi$ at every iteration), and $\mathbf{R}$ and $\bmath \lambda$ now include both the source and the potential corrections regularization.  

\subsection{Clumpy models}
\label{clumpy_model}
Once potential corrections have been identified, their significance can be quantified by comparing the marginalized Bayesian evidences of the analytical smooth models with that of an analytical clumpy model, i.e. a model made from the sum of a smooth host gravitational lens galaxy and an analytical mass substructure (for every potential correction that conservatively appear significant at a few $\sigma$ level). As in previous works (e.g. \citealt{Dalal02}), we assume the substructure to have a spherical pseudo-Jaffe profile, defined by the mass density,  
\begin{equation}
\rho(r) =  \frac{\rho_{0,{\rm{sub}}}}{r^2(r^2+r_{\rm t}^2)}\,,
\label{equ:rho_clump}
\end{equation}
and surface mass density,
\begin{equation}
\kappa(R) =  \frac{\kappa_{0,{\rm{sub}}}}{2}\left[R^{-1}-(R^2+r_{\rm t}^2)^{-1/2}\right]\,,
\label{equ:kappa_clump}
\end{equation}
\noindent where $\rho_{0,{\rm sub}} = \kappa_{0,{\rm{sub}}}\Sigma_c r_{\rm t}^2/2\pi$ and $r_{\rm t}$ is the half-mass radius.
 Assuming that the substructure is located in the plane of the sky that passes through the lens galaxy centre, we can interpret the truncation radius, $r_{\rm t}$, as a tidal radius and express it as a function of the substructure total mass $M_{\rm{sub}} = \pi\Sigma_c r_{\rm t}\kappa_{0,{\rm{sub}}}$. Under some approximations, the tidal radius can be expressed as
 \begin{equation}
r_{\rm t} = r\left(\frac{M_{\rm{sub}}}{ \beta M(<r)}\right)^{1/3}\,.
\label{equ:tidal}
\end{equation}
\noindent Here $M(r)$ is the 3D mass of the host galaxy within the spherical radius $r$ and $\beta$ is a factor which depends on the assumptions made, for example, on the orbit and size of the satellite and the contribution of centrifugal forces. If it is assumed that the gravitational potential of the satellite and the host are both given by point masses and that the satellite is small compared to its distance from the centre of the host, then $\beta=2$ (Roche limit). If the contribution of centrifugal forces  is included and if it is assumed that the satellite is on a circular orbit then $\beta=3$ (Jacobi limit). Finally for extended mass profiles, without the contribution of centrifugal forces and generic orbits, $\beta=2-d\ln M/d\ln r$ (Tormen et al 1998, $\beta=1$ for a SIS). 

For a pseudo-Jaffe substructure in a spherically symmetric isothermal galaxy, with mass normalization $\kappa_0$, the truncation radius in arcseconds is therefore $r_{\rm t} = r\sqrt{\pi\kappa_{0,{\rm{sub}}}/(2\beta\kappa_{0}})$, which reduces to $r_{\rm t}=\sqrt{\pi\kappa_{0,{\rm{sub}}}\kappa_{0}/(2\beta)}$ if the substructure is located at the Einstein radius of the lens. In addition to the free parameters defined in Section \ref{smooth_model}, the free parameters of the clumpy model now also include the substructure position and mass. 
In order to allow for a proper comparison, we use the same source regularization form and the same number of pixels $n_{\rm{src}}$ characterizing the source grid that we used for the smooth model analysis.  The posterior probability density distribution $P(\lambda_{\rm{s}},\bmath \eta\,|\,\bmath d,\mathbf{M_c},\mathbf R_{\rm{s}})$, is now maximized for $\bmath\eta=\{\kappa_0,\theta,q,x,y,\gamma,\Gamma,\Gamma_{\theta},M_{\rm{sub}},x_{\rm{sub}},y_{\rm{sub}}\}$ and $\lambda_{\rm{s}}$, following the same three-step procedure outlined above.
This analytic \emph{clumpy} model can also be used to assess the significance of the substructure detection via the total marginalized Bayesian evidence, as described in the following section. 

\subsection{Model comparison}
\label{model_comparison}
The marginalized Bayesian evidence is a measure of the probability of the data given the model, and provides an objective and quantitative way to compare and rank different models.
In our specific case, this marginalized evidence, ${\cal E}$, can be expressed as the integral of the normalization factor in equation (\ref{eq:evidence}) over the lens parameters $\bmath\eta$ and the source regularization $\lambda_{\rm{s}}$, such that
\begin{equation}
{\cal E}= P(\bmath d\,|\,\mathbf{M_c},\mathbf R_{\rm{s}})=\int{d \lambda_{\rm{s}}\, d\bmath
      \eta \,P(\bmath d\,|\,\lambda_{\rm{s}},\bmath \eta,\mathbf{M_c},\mathbf
      R_{\rm{s}}) P(\lambda_{\rm{s}},\bmath\eta)}\,.
    \label{equ:evidence_integral}
  \end{equation}
\noindent $P(\bmath d\,|\,\mathbf{M_c},\mathbf R_{\rm{s}})$ can then be turned into a probability density of the model itself using
  \begin{equation}
    P(\mathbf{M_c},\mathbf R_{\rm{s}}\,|\,\bmath d) \propto P(\bmath
    d\,|\,\mathbf{M_c},\mathbf R_{\rm{s}})P(\mathbf{M_c},\mathbf R_{\rm{s}})\,.
 \end{equation}
\noindent The integral of equation (\ref{equ:evidence_integral}) is calculated numerically with the {MultiNest} method by \citet{Feroz08}, which is based on the original nested sampling idea by \citet{Skilling04}. 
The method requires a defined prior probability density distribution $P(\lambda_{\rm{s}},\bmath\eta)$ for each of the parameters. 
Given our limited knowledge, we assume uniform prior distributions on $\bmath \eta$ and on $\log{\lambda_{\rm{s}}}$. These priors are the same for both the smooth and the clumpy models.
For the substructure mass and position in the clumpy models, we assume a uniform prior between $4 \times10^6M_\odot$ and $4\times10^{10}M_\odot$, and a uniform prior inside the lensed image grid, respectively.
Note that in the analysis presented in Section \ref{mass_function}, we effectively update the prior on the mass by including both the sensitivity function and the substructure mass function.

\section{Lens modelling results}
In this section, we provide more detailed comments on the individual systems and give comparisons with previous results where appropriate.

\subsection{Smooth models}
\begin{figure*}
\begin{center} 
\includegraphics[width = 14 cm]{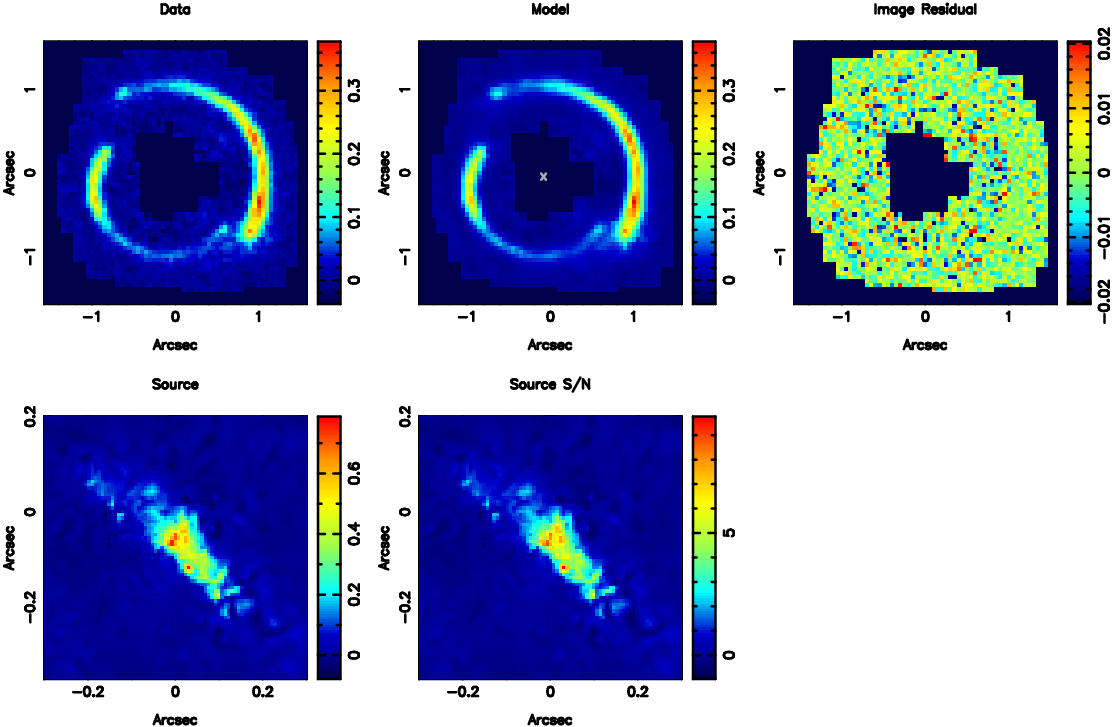}
\caption{Best smooth models for the lens gravitational system J0252+0039 observed with HST in the F814W filer.}
\label{fig:J0252_smooth}
\end{center}     
 \end{figure*}

\begin{figure*}
\begin{center} 
\subfigure{
\includegraphics[width=14 cm]{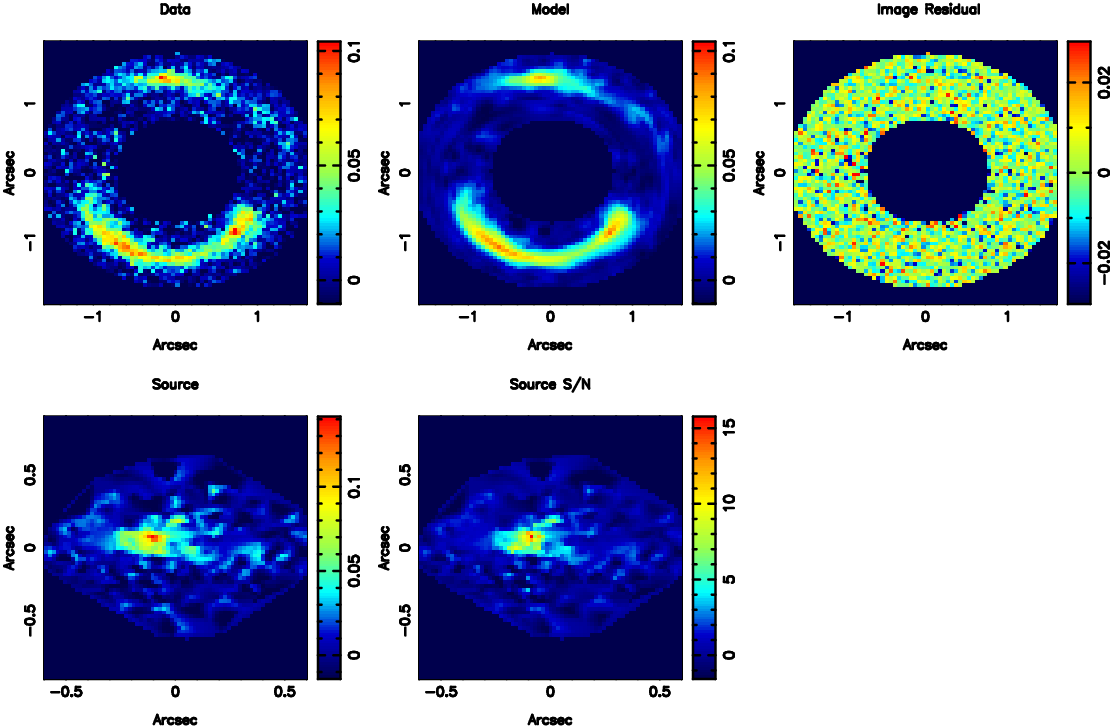}}
\subfigure{
\includegraphics[width=14 cm]{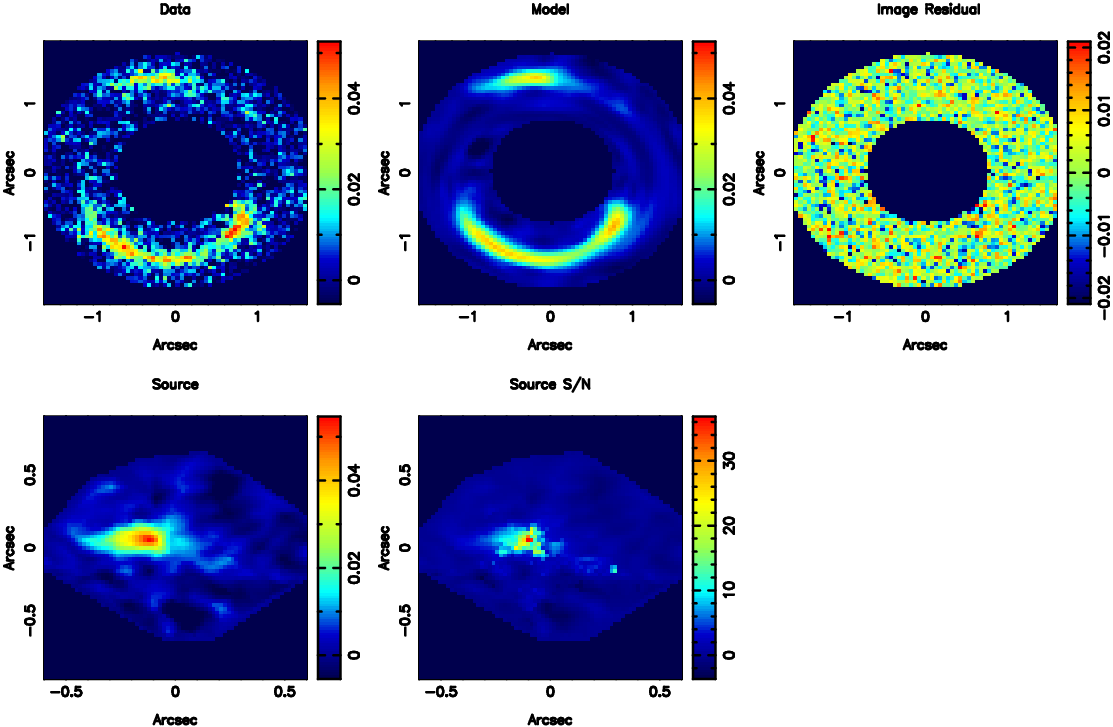}}
\caption{Same as figure \ref{fig:J0252_smooth} for J2300+0022 as observed in the HST F814W filter (top) and in the HST F555W filter (bottom).}
\label{fig:J2300_smooth}
\end{center}     
\end{figure*}

The MAP models under the assumption of a smooth mass distribution are listed in Table \ref{tbl:smooth_modeling_results} for each system and plotted in Figs. \ref{fig:J0252_smooth} to \ref{fig:J2300_smooth}.
Models for SDSS\,J0946+1006 can be found in \citet{Vegetti10}.
Smooth models for all of the SLACS lenses, under the assumption of a singular isothermal elliptical (SIE) mass distribution, were derived by \citet{Bolton08a} \citep[see also][]{Auger09}, while for a subset of the lenses considered here, a joint lensing and 2D kinematics analysis can be found in the paper by \citet{Barnabe11}. A measure of the slope of the mass density profile $\gamma$ was also derived by  \citet{Koopmans09} and \citet{Auger10}, using a combination of lensing and the stellar velocity dispersion for each galaxy. We note that the comparison between our results and those by \citet{Koopmans09}, \citet{Bolton08a} and \citet{Barnabe11} is not a straightforward one and that agreement should not be necessarily expected. 
For example, the mass density slope derived in this paper is a measure of the local slope near the Einstein radius and could be therefore different from the average mass density slope measured by \citet{Koopmans09} and \citet{Barnabe11}, which are mostly set to fit the dynamical data \citep[see discussion by][]{Sonnenfeld13}.  
We do find that the Einstein radius derived in this paper is for every lens consistent with those determined by \citet{Bolton08a}, although our lenses are often not (but close to) isothermal.

The lens system SDSS\,J1430+4105 was modelled by \citet{Eichner12} by making use of different assumptions on the mass model for the main lens and its environment, and providing consistent results within the error among several models. 
Here, we can only directly compare with the power-law plus external shear model from \citet{Eichner12} derived by modelling the surface brightness distribution of the lensed images with the lensing code { \sc LENSVIEW} \citep{Wayth06}. We find that the Einstein radius derived in this paper is $\sim$1 percent smaller than their Einstein radius, while our mass distribution has an axis ratio which is  $\sim$15 percent smaller and a mass density slope which is $\sim$10 percent steeper.
Finally, the external shear strength is $\sim$8 percent larger than the value recovered by \citet{Eichner12} and its position angle is not consistent with their model, but it is consistent with the location of the brightest cluster galaxy of the lens environment \citep{Koester07}. These differences are consistent with the scatter on the lens parameters derived by \citet{Eichner12} under different model assumptions. We believe that they are related to a different choice of the surface mass density normalization, which in our case explicitly encodes the mass distribution axis ratio and logarithmic slope (equation \ref{equ:kappa_smooth}). Also, the power-law plus external shear model by \citet{Eichner12} does not seem to recover the external shear correctly.

In the case of SDSS\,J0956+5100 a luminous companion was identified in the {\it HST} images. This luminous satellite was included in the lens model as an additional mass component with a truncated pseudo-Jaffe (see equation \ref{equ:kappa_clump}) mass density profile. 

\begin{table*}
\caption{The best-fitting (i.e. MAP) parameters for our smooth reconstruction lens models. 
  The mass-model parameters are defined as follows:  $\kappa_0$ is the normalization, $\theta$ the position angle, $q$ the axis ratio and $\gamma$ the power-law slope.
  $\Gamma$ and $\Gamma_{\theta}$ are, respectively, the magnitude and the position angle of an external shear component. The number $n_{\rm{src}}$ of pixels characterizing the source grid and the form of source regularization are given in columns 9 and 10, respectively. $C$ is the curvature regularization and $G$ is the gradient regularization; the subscript adp is used to indicate cases where the source regularization is adaptive with the lensed images S/N. $^a$Total mass of the luminous satellite in units of $10^{10}M_\odot$, for a pseudo-Jaffe profile.}
\begin{tabular}{cccccccccc}
\hline
 Name (SDSS) &Filter&$\kappa_0$  &$\theta$ [deg.]  &$q$  &$\gamma$  &$\Gamma$  &$\Gamma_{\theta}$ [deg.]&$n_{\rm{src}}$&$\mathbf {R_{\rm{s}}}$\\
 \hline

$\rm{J0252+0039}$&$I$&1.022&116.2&0.943&2.047&0.009&101.8&1&$G_{\rm{adp}}$\\

$\rm{J0737+3216}$&$I$&0.951& 98.31&0.705&2.066&0.050&100.8&1&$C$\\
 
 &$V$&0.951&97.18&0.709&2.073&0.052&102.5&2&$G_{\rm{adp}}$\\ 

$\rm{J0956+5100}$&$I$&1.260&144.9&0.656&2.060&0.037&140.3&2&$C$\\
&&1.68$^a$\\

&$V$&1.253&145.6&0.651&2.062& 0.035&139.0&2&$C$\\
&&1.93$^a$\\

$\rm{J0959+4416}$&$I$&0.902&95.06&0.955&2.060&0.033&65.67&2&$C_{\rm{adp}}$\\

$\rm{J1023+4230}$&$I$&1.329&157.9&0.926&2.286&0.050&174.4&2&$C_{\rm{adp}}$\\

$\rm{J1205+4910}$&$I$&1.199&153.0&0.706&2.047&0.025&172.7&1&$C_{\rm{adp}}$\\

&$V$&1.197&152.9&0.706&2.044&0.027&173.6&1&$C_{\rm{adp}}$\\ 

$\rm{J1430+4105}$&$I$&1.484&106.5&0.710&2.048&0.051&128.6&1&$C$\\

$\rm{J1627-0053}$&$I$&1.229&9.260&0.912&1.998&0.004&79.96&2&$C_{\rm{adp}}$\\

&$V$&1.212&9.202&0.869&2.058&0.014&87.76&2&$C_{\rm{adp}}$\\ 

$\rm{J2238-0754}$&$I$&1.241&139.9&0.786&2.121&0.009&78.40&1&$G_{\rm{adp}}$\\

&$V$&1.243&139.2&0.780&2.117&0.009&80.93&2&$C_{\rm{adp}}$\\

$\rm{J2300+0022}$&$I$&1.196&77.49&0.694&2.131&0.048&109.4&2&$C$\\

&$V$&1.201&84.09&0.687&2.130&0.042&102.8&2&$C$\\ 

 \hline
\end{tabular}

\label{tbl:smooth_modeling_results} 
\end{table*}

\subsection{Potential corrections and clumpy models}
\label{sec:clumpy_res}
We claim the presence of a substructure detected at a significant statistical level when the following conditions are simultaneously satisfied:
\begin{enumerate}
\item a positive convergence correction that improves the image residuals is found independently from the potential regularization, number of source pixels, PSF rotations, and galaxy subtraction procedure;\\
\item a clumpy model is preferred over a smooth model with a Bayes factor $\Delta\log{{\cal E}}=\log{{\cal E}_{\rm{smooth}}}-\log{{\cal E}_{\rm{clumpy}}}\ge-50$ (to first order equivalent to a 10$\sigma$ detection, under the assumption of Gaussian noise);\\
\item the mass and the position of the substructure obtained via the nested sampling analysis are consistent with those independently obtained by the potential corrections and the MAP parametric clumpy model;\\
\item the results are consistent among the different {\it HST} filters, where available.
\end{enumerate}
We find that none of the systems considered in this paper satisfies the above conditions, implying that no significant detection of a mass substructure was made above the mass threshold set by the S/N and resolution of the {\it HST} images. Importantly, we notice that some lenses satisfy the condition (ii) without satisfying conditions (iii) and (iv). We believe that this is due partly to degeneracies in the lens model parameters, a relatively poor fitting of the smooth model and the fact
that only a clumpy model was considered as an alternative to the smooth one. Generalizing, we stress that is important to verify the uniqueness of the substructure detection by taking fully into account the degeneracies with the macro model and the source model and by testing other models than clumpy ones as an alternative to smooth models, both with gravitational imaging and with flux ratio anomalies.

\subsubsection{SDSS\,J0956+5100}
For the case of SDSS\,J0956+5100, since the luminous satellite was explicitly included as part of the smooth model, this implies that no substructure other than the luminous satellite was found. By repeating the analysis of this lens without the inclusion of the luminous satellite, we find that this satellite is also detected via the gravitational imaging technique in both $I$ and $V$ bands at the $\sim$37 and $\sim$20$\sigma$ level, respectively. Including our previous detection in SDSS\,J0946+1006 \citep{Vegetti10}, we detected in total, two satellites in 11 gravitational lens systems. In both cases the satellites have a projected distance of about $\le 5$ kpc from the centre of the host galaxy. Although we do not know its redshift, the colours of this luminous clump are consistent with those of the lens galaxy, indicating that it is most likely a satellite.
By fitting the surface brightness of this satellite with a Gaussian profile, we find that the total magnitude is 22.70 and 22.93 in $I$- and $V$-band, respectively, which is about 6 mag fainter than the host galaxy. From \citet{Auger09} we derive a stellar mass of $M_* = 1.4\times 10^9 M_\odot$ for a Chabrier initial mass function and $M_* = 2.5\times10^{9}M_\odot$ for a Salpeter one. 
Under the assumption of a pseudo-Jaffe total mass density profile, we infer a total mass of $M_{\rm{sub}}=(1.69\pm0.01)\times10^{10}M_\odot$ and $M_{\rm{sub}}=(1.89\pm0.05)\times10^{10}M_\odot$ in $I$ and $V$ band, respectively. 
This yields a stellar- to total-mass ratio of $\sim$10, which is not unexpected considering that the satellite has probably been significantly tidally stripped. 
Finally, we note that the MAP position is located in both bands offset from the centre of the light distribution by 3 pixels (0.15 arcsec) in the $x$ direction and 8 pixels (0.4 arcsec) in the $y$ direction.

\subsubsection{SDSS\,J0946+1006}
\citet{Vegetti10} reported the detection of a dark-matter-dominated substructure in the gravitational lens system SDSS\,J0946+1006.
When modelled as a pseudo-Jaffe profile, the total mass of this substructure was inferred to be $M_{\rm{sub}} = (3.51 \pm 0.15) \times 10^9 M_\odot$.
The substructure was located very close to the Einstein radius but in a position where the surface brightness is not dominated by the lensed images. This allowed a tight constraint to be set on the upper limit of the substructure luminosity, hence its mass-to-light ratio lower limit is $(M/L)_{V,\odot}\ge 120~M_\odot/ L_{V,\odot}$ (3-$\sigma$) inside a sphere of radius 0.3 kpc.\\

\subsubsection{Systematic error on the substructure masses}
\label{systematic_mass}
The major source of systematic error on the substructure total mass is related to the de-projection of the substructure position. In particular, we have assumed that the detected mass substructure and satellites are at the same redshift as the lens galaxy, and when using a pseudo-Jaffe profile, that the observed projected position is the true physical position.  As discussed previously, in the case of SDSS\,J0956+5100,  the luminous satellite has colours consistent with the lens galaxy redshift. The mass clump in SDSS\,J0946+1006 is a high mass-to-light ratio object for which direct light was not detected and colours could not be measured. In principle, we can estimate the probability that this clump is at the same redshift of the lens galaxy by making use of numerical simulations, in practice however, current $N$-body simulations of massive galaxies do not reach the necessary resolution, especially at distances so close to the centre of the host halo. Moreover, one would need to take into account for the different lensing effects of clumps at different redshifts. This can significantly depend on the density profile that is assumed and, in the case of extended lensed images, it cannot be computed by linearly rescaling the clump mass. These are all important issues that we will extensively address in a future paper. 

Here we limit ourselves to the assumption that the clump detections are indeed substructure and quantify the systematic error on their mass due to de-projection within the host galaxy, under the assumption of an isothermal spherically symmetric host galaxy. Making use of Bayes theorem we can express the probability density of the substructure position $r$ in 3D, given an observed projected position $R$ as \citep{Suyu10}
\begin{equation}
P(r|R) = \frac{P(R|r)P(r)}{P(R)} = \frac{1}{r\sqrt{r^2/R^2-1}~\arccos[R/r_{\rm{max}}]}\,.
\end{equation}
\noindent Following the results by \citet{Nierenberg11, Nierenberg12}, we assume that the 3D spatial distribution of the satellites follows the total mass distribution of the host galaxy. To normalize $P(r)$, we impose that the total mass density profile of the host galaxy goes to zero at an arbitrarily large radius well beyond the lensed images  (e.g. $r>r_{\rm{max}} \sim100$ arcsec) and at $r<R$. For a pseudo-Jaffe profile, we can express the total mass as a function of the position $r$ by combining the expression of the total mass $M_{\rm{sub}}=\pi\Sigma_c r_{\rm t}\kappa_{0,\rm{{sub}}}$ with the expression for the tidal radius (equation \ref{equ:tidal}). From this we derive the probability density for a given substructure mass 
\begin{equation}
P(M_{\rm{sub}}|R) =\frac{1}{M_{\rm{sub}}\sqrt{\frac{2M_{\rm{sub}}^2\beta\kappa_0}{R^2\kappa_{\rm{sub}}^3\pi^3\Sigma_{\rm c}^2}-1}\arccos[R/r_{\rm{max}}]}\,,
\label{equ:m_r_prob}
\end{equation}
\noindent for $R\Sigma_{\rm c}\sqrt{\frac{(\pi\kappa_{\rm{sub}})^3}{2\beta\kappa_0}}\le M_{\rm{sub}}\le r_{\rm{max}} \Sigma_{\rm c}\sqrt{\frac{(\pi\kappa_{\rm{sub}})^3}{2\beta\kappa_{\rm 0}}}$ and zero otherwise. This leads to a probability density for the substructure mass that is strongly peaked at the value measured under the assumption $r\equiv R$. We derive that de-projection effects lead to an error on the logarithmic substructure mass of $\sigma_{M_{\rm{sub}}}=^{+1.17}_{-0.17}$ and $\sigma_{M_{\rm{sub}}}=^{+1.11}_{-0.16}$ at the 68 percent CL for SDSS\,J0946+1006 and for SDSS\,J0956+5100, respectively. These uncertainties can be quite large since the extrapolation from the lensing mass to the total mass significantly increases with the distance of the substructure from the plane of the lens; nevertheless, masses within small apertures (e.g. 300 pc) are still precisely determined.

\section{Substructure mass function}
\label{mass_function}
In this section, we use the results of this paper, in combination with the mass substructure detected by \citet{Vegetti10}, to constrain the substructure mass function slope and normalization. For this purpose, we assume that the detected mass density corrections are due to mass clumps associated with the lens galaxy, and not with line-of-sight contaminations. In the latter case, the inferred substructure fractions in the galaxy would decrease. Although only one out of 11 lens galaxies shows evidence for mass substructure, the remaining 10 non-detections also provide valuable information on the mass function, once the sensitivity to substructure for each lens galaxy has been quantified. For each lens in the considered sample, we quantify the substructure sensitivity in the following subsection.

\subsection{Substructure sensitivity function}
\label{sensitivity}
\citet{Vegetti09b} and \citet{Vegetti10, Vegetti12} assumed that the sensitivity to substructure was constant (i.e. it only depended on the substructure mass but not its position) over an annulus of a few arcseconds width around the Einstein radius and that substructures were randomly distributed within this annulus with a uniform probability density distribution. In practice, this means assuming a constant value for the lowest detectable substructure mass within the considered region. In this paper, we quantify the sensitivity to substructure for each lens system in the considered sample as a function of the substructure position and mass. This is particularly important to properly take into account for the large number of non-detections.

\begin{figure}
\includegraphics[width=0.48\textwidth]{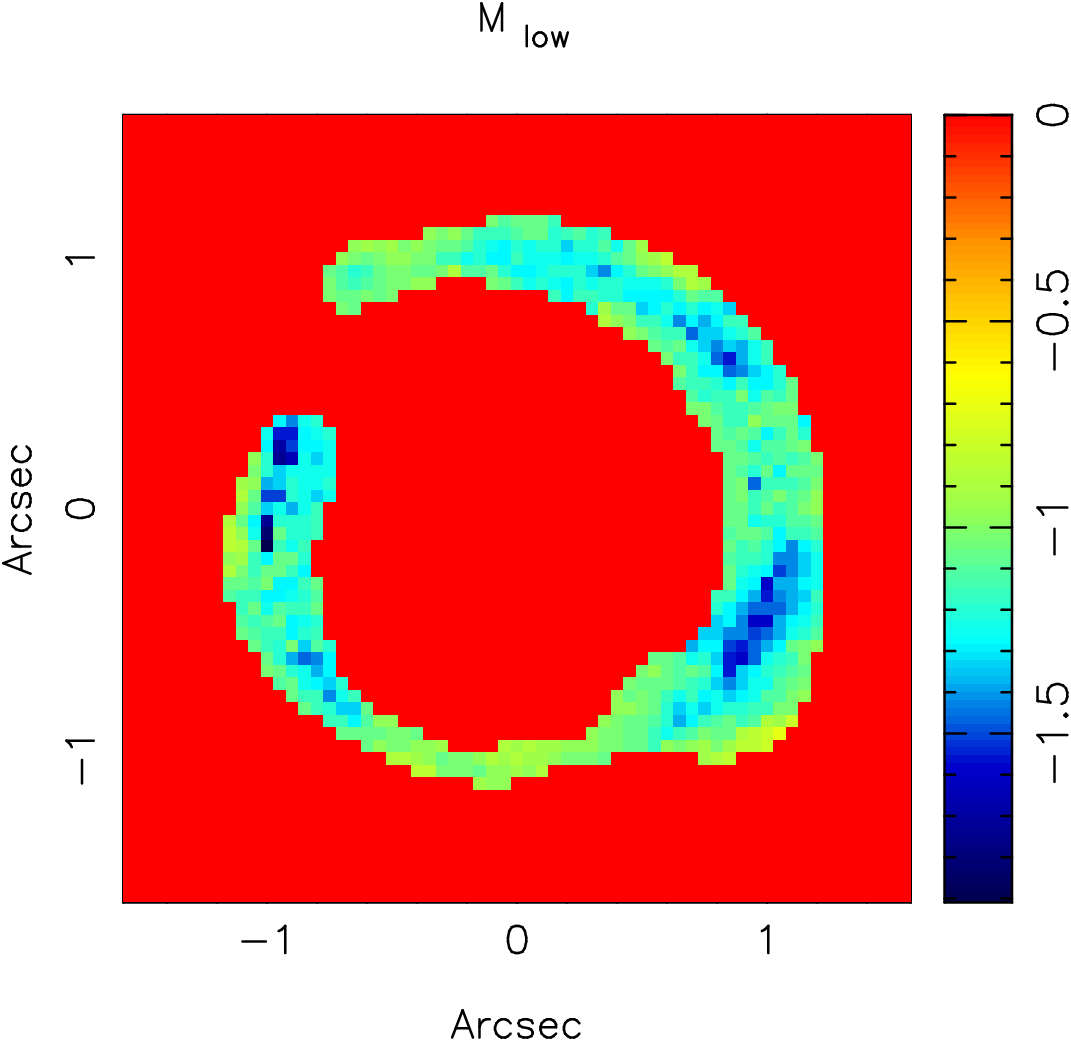}
\caption{Lowest detectable mass (in $\log[M_{\rm{low}}/ 10^{10}M_\odot]$) as function of the substructure position for the lens system SDSS\,J0252+0039.}
\label{fig:sens_func}
\end{figure}

As shown by \citet{Koopmans05} and \citet{Vegetti09a}, the effect of a potential correction on the surface brightness distribution of the lensed images is given by $\delta I(\mathbf{x})=\nabla S(\mathbf{y})\cdot\nabla\delta\psi(\mathbf{x})$, where $\delta I(\mathbf{x})$ is the difference in the surface brightness distribution of the lensed images between a smooth lens and a model that contains substructure, $\nabla S(\mathbf{y})$ is the gradient of the source surface brightness distribution and $\nabla\delta\psi(\mathbf{x})$ is the gradient of the potential correction $\delta\psi(\mathbf{x})$. For a given substructure, therefore, its effect is different on different parts of the lensed images and larger on those parts which are more structured \citep[see][]{Rau13}. This, in combination with the fact that the effect of a substructure can be partly re-absorbed with a change of the source surface brightness structure and/or with a change in the lens macro model, implies that the most rigorous way to quantify the sensitivity to a given substructure is via the probability density of a specific substructure model given the data, marginalized over all possible macro model parameters and source regularization constant (i.e. via equation \ref{equ:evidence_integral}). This would require one to compute the Bayes factor and marginalize over the lens parameters and source parameters for a large number of clumpy models, each defined by a different combination of substructure mass and position (ideally one for every image pixel). However, this approach is computationally prohibitive, and the {\sc MultiNest} technique does not  sample the posterior probability density distribution densely enough to allow us to derive the sensitivity function directly from the analysis of Section \ref{model_comparison}. At the same time, we find that most of the re-absorption is done by changes in the source surface brightness distribution rather than changes in the macro-model parameters. 
We can therefore reduce the multidimensional integral of the  Bayesian evidence to a one-dimensional integral over the source regularization constant, while keeping the lens (and substructure) parameters fixed. Because this integral implicitly involves also an integral over the source surface brightness distribution via equation (\ref{equ:posterior_2}), it properly takes into account for the degeneracy between the substructure properties and the structure of the source. We, therefore, properly quantify the sensitivity function by creating $100\times N_{\rm{pix}}$ mocked lens realizations, each of which is based on the most probable smooth lens model and most probable source model as derived in Section \ref{smooth_model}, and has a substructure of given mass between $4\times10^6 M_\odot$ and $4\times 10^{10} M_\odot$ (uniformly sampled in log-scale) and located at one of the image pixels. This assumes that we are not sensitive to changes in the substructure position within a single pixel.  For each of these realizations, we then compute the marginalized Bayesian evidence under the assumption of a smooth and a clumpy lens model and define the mass detection threshold at each position on the lens plane as the lowest mass for which the Bayes factor is $\Delta\log{{\cal E}}\ge-50$ (see Fig. \ref{fig:sens_func}), as this would correspond to a 10$\sigma$ detection, under the assumption of Gaussian noise.  We find that given the data quality of the sample in this paper, the lowest detectable mass typically varies on average between $0.04\times10^{10}M_\odot$ and $0.14\times10^{10}M_\odot$ depending on the lens. In some cases, for specific substructure positions the detection threshold, however, can be as low as $0.01\times10^{10}M_\odot$.

{Because this approach is still relatively computationally expensive, one may be tempted to quantify the effect of substructures by using the difference in the lensed images surface brightness distribution (i.e. the image residuals) of different clumpy models relatively to a smooth model via the following $\chi^2$ instead, 
\begin{equation}
{\cal D} = \sum_x{\left(\frac{I_{\rm {smooth}}(x)-I_{\rm {clumpy}}(x)}{\sigma(x)}\right)^2}\,.
\end{equation}
Using mock data realizations one could then translate a detection threshold on ${\cal D}$ into a substructure mass $M_{\rm{low}}(\vect{x})$ detectable at the 10$\sigma$ level, as a function of the lensed images and substructure position. 
However, we caution that this simplistic approach does not take into account for the degeneracies described above and that may lead to an overestimation of the sensitivity function.
In order to highlight the approximation of this approach,} we calculate ${\cal D}$ for the same $100\times N_{\rm{pix}}$ mocked lens realizations described above under the assumption of a smooth and clumpy model. We find that although the likelihood ratio is not significantly affected by changes in the lens parameters and source surface brightness distribution, the two models are practically indistinguishable in terms of their posterior. This clearly indicates that \emph{re-absorption} of the substructure effect  by changing the source and the lens model has a significant influence on the significance of a substructure detection. In the light of these considerations, we caution therefore that substructure detections based only  on likelihood ratio tests or equivalently on $\chi^2$ difference criteria should not be necessarily believable. Moreover, in cases with a large number of non-detections, such as the one considered in this paper, this can significantly bias constraints on the mass function parameters. We find that the lowest detectable substructure mass as properly derived above can be up to two orders of magnitude larger than the limit obtained using equation (14), due to the fact that the effects of such low-mass structures can quite easily be absorbed in a small change to the source model.

Finally, we note that re-absorption effects were properly taken into account in the detection of the mass substructure in the lens system SDSS\,J0946+1006, as its significance was properly calculated via the integral in equation (\ref{equ:evidence_integral}). We refer to \citet{Vegetti10} for a more detailed discussion on this particular lens systems.
\subsection{Likelihood of substructure detections}
A statistical formalism to turn $n_{\rm{s}}$ substructure detections (each with mass $m_{\rm{s}}$ and position $R_{\rm{s}}$) and non-detections into constraints on the substructure mass function was derived by \citet{Vegetti09b}. In particular, a posterior probability density distribution $P\left( \alpha,f ~|~ \{n_{\rm{s}},\vect{m_{\rm{s}},R_{\rm{s}}}\},\vect{p} \right)$ for the mass function slope $\alpha$ and normalization $f$ (i.e the projected fraction of dark matter mass in substructure within the considered region around the Einstein radius, see below for a definition of this region) was derived. $P\left( \alpha,f ~|~ \{n_{\rm{s}},\vect{m_{\rm{s}},R_{\rm{s}}}\},\vect{p} \right)$ is related, via the Bayes theorem, to the likelihood ${\cal L}\left( \{n_{\rm s},\vect{m_{\rm{s}},R_{\rm{s}}}\}\right)$ of detecting a certain number $n_{\rm s}$ of substructures of given masses $\vect{m}$ and to the prior probability density distribution on both $\alpha$ and $f$. This approach also requires the cumulative projected dark matter mass of the host lens galaxy to be known. 
We extend the formalism presented by \citet{Vegetti09b} for the case of a substructure mass detection threshold that changes with the considered lens system and with the  position of the substructure on the lens plane. By assuming that the number of substructures has a Poisson distribution, the following expression for the likelihood of detections ${\cal L}\left( \{n_{\rm{s}},\vect{m_{\rm{s}},R_{\rm{s}}}\}\right)$ can be derived for one lens, $n_{\rm{s}}$ detections and model parameters $\vect{p}=\{M_{\rm{min}},M_{\rm{max}},M_{\rm{low}}\}$ (minimum substructure mass, maximum substructure mass, lowest detectable substructure mass):
\begin{multline}
{\cal L}\left(\{n_{\rm{s}},\vect{m_{\rm{s}},R_{\rm{s}}}\}~|~\alpha,f(<R),\vect{p}\right) =\\ \frac{e^{-\mu(\alpha,f,\vect{p})}~{\mu(\alpha,f,\vect{p})}^{n_{\rm{s}}}}{n_{\rm{s}}!}\prod_{k=1}^{n_{\rm{s}}}P\left(m_{k},R_{k}~|~\vect{p},\alpha\right) \,.
\label{equ:single_lens}
\end{multline}
\noindent Here, $\mu(\alpha,f(<R),\vect{p}) =\sum_{j=1}^{{N_{\rm{pix}}}}\mu_{j}(\alpha,f,\vect{p})$ is the total expectation of substructures with masses larger than the detection threshold summed over the number of pixels on the lens plane within the considered region, and $\mu_{j}(\alpha,f,\vect{p})$ is given by
\begin{multline}
\mu_{j}(\alpha,f,\vect{p})= \mu_{0,j}(\alpha,f,\vect{p})\int_{M_{\rm{low,j}}}^{M_{\rm{max}}}P\left(m,R_{j}~|~\vect{p},\alpha\right) dm\\
=  \mu_{0,j}(\alpha,f,\vect{p})\int_{M_{\rm{low,k}}}^{M_{\rm{max}}}\frac{dP}{dm}dm\,	
\label{equ:mu}	
\end{multline}
\noindent for
\begin{equation}
\left\{
 \begin{array}{cc}
\mu_{0,j}(\alpha,f,\vect{p}) = \frac{f_{j} M_{\rm{DM}}(<R)}{ \int_{M_{\rm{min}}}^{M_{\rm{max}}}{m~\frac{dP}{dm}~dm}}\\
	\\
f_{j} = \frac{f(<R)}{N_{\rm{pix}}}	
\end{array}
\right.	
\label{equ:mu_not}	
\end{equation}
\noindent and
\begin{equation}
\frac{dP}{dm}=\left\{
 \begin{array}{cc}
 \frac{\left(1-\alpha\right)~m^{-\alpha}}{ \left(M_{\rm{max}}^{1-\alpha}~-~M_{\rm{min}}^{1 -\alpha}\right)}& \alpha \neq 1\\
\\
 \frac{m^{-\alpha}}{ \ln{\left(M_{\rm{max}}/M_{\rm{min}}\right)}}& \alpha = 1 
\end{array}
	\,.\right.
\end{equation}
 For each detected substructure, $P\left(m_k,R_{k}~|~\vect{p},\alpha\right)$ is the probability density of observing one substructure with mass within the Einstein radius $m_k$ and observed projected position $R_k$. This probability density can be derived from the unknown 3D total mass $m$ and position $r$ by assuming a 3D spatial distribution of the substructure, $P(r)$ and related projection effects via $P(R|r)$. As in Section \ref{systematic_mass}, we assume the substructure spatial distribution to follow the isothermal host galaxy total mass density distribution (see \citealt{Nierenberg11, Nierenberg12,Wang14} and references therein for an overview on observational and theoretical constraints) and obtain: 
\begin{multline}
P\left(m_k,R_{k}~|~\vect{p},\alpha\right) = \\ \frac{\int_{M_{\rm{low,k}}}^{M_{\rm{max}}}\int_{R_k}^{r_{\rm{max}}}{\cal N}(m_{k},\sigma_{m_k}| m_e)m^{-\alpha}P(R_k|r)P(r)~dm~dr} { \int_{M_{\rm{min}}}^{M_{\rm{max}}}\int_{R_k}^{r_{\rm{max}}}{\cal N}(m_{k},\sigma_{m_k}| m_e)m^{-\alpha}P(R_k|r)P(r)~dm~ dr}\,,
\end{multline}
where $m_e$ is the predicted mass within the Einstein radius and is a function of the total 3D mass and the 3D position $r$.
It is important to note, that we are not interested here in constraining the mass function of each single lens galaxy and that the fraction $f$ and the slope $\alpha$ are therefore defined at the galaxy population level instead. 
In practice, this means that the galaxy/subhalo systems are assumed to be self-similar and all following the same mass function. Since we are considering a sample of 11 galaxies with a narrow range in mass and redshift, we can consider the above assumption valid. In particular, $f$ is defined as the mean projected mass fraction for substructure masses between $M_{\rm{min}} = 4 \times 10^6M_\odot$ and $M_{\rm{max}} = 4\times 10^9M_\odot$.

\subsubsection{Considered region}
Given a substructure mass, its effect on the surface brightness of the lensed images is a function of the substructure distance from the lensed images, the data angular resolution and S/N, and the level of structure of the source surface brightness distribution. Thus, for a given data quality (S/N and angular resolution) the gravitational effect of a mass substructure is maximized when located right on top of the lensed images of a highly structured source. 
We therefore restrict our measurement of the mass function to a small region where the S/N of the data surface brightness distribution is larger than 3. This approximately corresponds to a small area around the critical curve that can vary between $\sim$4 and $\sim$200 kpc$^2$ (i.e. between 2 and 13 arcsec$^2$) around the lensed images. In this way, we are only considering substructure positions where the effect of the substructure is large and unlikely to be re-absorbed by changes in the lens macro model. We note that this region does not include the luminous satellite of SDSS\,J0956+5100, after having subtracted the light from the satellite itself, hence this luminous satellite is not included in the Bayesian inference of the substructure mass function.

Our mass function constraints are therefore solely derived by making use of the substructure detected in SDSS\,J0946+1006 and the 10 other non-detections.
We note that changes in the window in which we allow a detection does not bias the inference since we properly account for the mass detection threshold and dark matter mass inside the window. Shrinking the window, however, does lower the detection probability, increasing the error bars. But, since the expectation value of the number of substructures (per unit area) strongly peaks around the lensed images, the precise window choice (if not too small) has little impact and choosing it relatively narrow significantly reduces the computation effort in deriving the mass thresholds as a function of position in the image plane.

\subsubsection{Lens galaxy dark matter mass}
\citet{Vegetti09b} and \citet{Vegetti10, Vegetti12} calculated the host lens galaxy dark matter mass assuming an isothermal total mass density distribution, a Jaffe profile for the stellar mass distribution and the scaling relations derived for the SLACS lenses by \citet{Bolton08b}. Here, we make use instead of the dark matter fraction derived by \citet{Auger10}. In particular, we evaluate the de Vaucoleurs surface brightness profiles from \citet{Bolton08b} and \citet{Auger09} at the Einstein radius and then multiply by the mass-to-light ratio implied by \citet{Auger09} under the assumption of a Salpeter initial mass function \citep [consistent with results from lensing and dynamics, stellar population models and stellar kinematics; e.g.][]{Treu10, Auger10,Spiniello11, Cappellari12, Barnabe13}. The dark matter mass within the considered region is then calculated by subtracting the stellar mass so derived from the total mass measured with the gravitational lens modelling.

\begin{table}
\begin{center}
\caption{Mass function constraints: substructure and mass function normalization and slope for a uniform (U) and a Gaussian (G) prior on the slope, with relative marginalized evidence in the last column.}
\begin{tabular}{cccc}
\hline
$P(\alpha)$&$f$ (68\% CL)&$\alpha$ &$ \ln{\rm{Ev}}$\\
 \hline
 U&\ufr&\ual ~(95\% CL)&\Evu\\
 \\
 G&\gfr&\gal ~(68\% CL) &\Evg\\
\hline
\end{tabular}
\label{tbl:clumpy_modeling_results}
\end{center}
\end{table}
 
\begin{figure}
\includegraphics[width =0.48\textwidth]{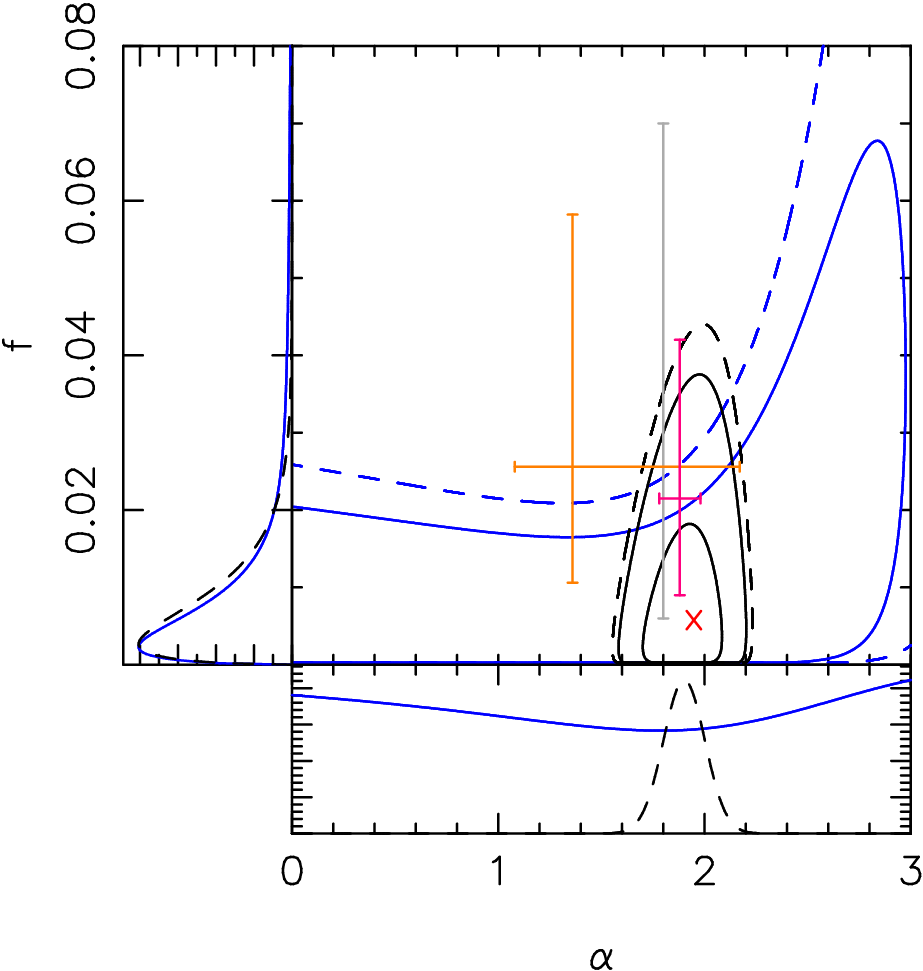}
\caption{Joint probability $P (\alpha, f | {n_{\rm{s}} , m}, p)$ contours and marginalized probabilities $P( f | {n_{\rm{s}} , m}, p)$ and $P (\alpha | {n_{\rm{s}} , m}, p)$ assuming a uniform prior (blue lines) and assuming a Gaussian prior in $\alpha$ (black lines). Contours (inside out) are set at levels $\Delta \ln(P) = -1, -4, -9$ from the peak of the posterior probability density. The cross represents the prediction from the CDM model as derived by \citet{Xu13}, the solid grey line is the range of mass fractions allowed by flux-ratio anomalies in the multiply imaged quasar sample analysed by \citet{Dalal02}, while the results derived by \citet{Vegetti10} from the single detection in the lens system SDSS\,J0946+1006 with a uniform prior and with a Gaussian prior on the slope $\alpha$ are shown by the orange and magenta points respectively.}
\label{fig:mass_function}
 \end{figure}
 
\subsection{Posterior probability of $\alpha$ and $f$}
\label{sec:posterior_mf}
The Bayes theorem can then be used to derive a joint posterior probability density function for the mass function slope and normalization:
\begin{multline}
P\left( \alpha,f ~|~ \{n_{\rm{s}},\vect{m_{\rm{s}},R_{\rm{s}}}\},\vect{p} \right) =\\
\frac{{\cal L}\left( \{n_{\rm{s}},\vect{m_{\rm{s}},R_{\rm{s}}}\}~|~\alpha,f,\vect{p}\right)P\left( \alpha,f ~| ~\vect{p}\right)}{P\left( \{n_{\rm{s}},\vect{m_{\rm{s}},R_{\rm{s}}}\}~|~\vect{p}\right)}\,,
\end{multline}
\noindent where $P\left( \alpha,f ~| ~\vect{p}\right)= P\left( \alpha ~| ~\vect{p}\right)P\left(f ~| ~\vect{p}\right)$ is the prior probability density on the relevant parameters. 
We assume the normalization $f$ to have a non-informative Jeffreys' prior probability density distribution \citep{Jeffreys46} between $f_{\rm{min}}=0$ and $f_{\rm{max}}=1$:
\begin{equation}
P\left(f \right) = \frac{1}{2\left(\sqrt{f_{\rm{max}}}-\sqrt{f_{\rm{min}}}\right)\sqrt{f}}\,.
\end{equation}
Two different priors are considered for the slope $\alpha$ instead; in the first case we assume a non-informative uniform prior density distribution, between
$\alpha_{\rm{min}}=0$ and $\alpha_{\rm{max}}=3$, in the second case we assume a Gaussian prior centred on $\alpha_{\rm{mean}}=1.9$ and with a full width at half-maximum (FWHM) of  $\sigma_{\alpha}=0.1$, as suggested by $N$-body simulations \citep{Springel08, Xu13}:
\begin{equation}
P_{\rm{U}}\left(\alpha\right) = \frac{1}{\alpha_{\rm{max}}-\alpha_{\rm{min}}}\,
\end{equation}
and
\begin{equation}
P_{\rm{G}}\left(\alpha ~| ~\vect{p}\right) =\frac{1}{\sigma_{\alpha}\sqrt{2\pi}}\exp{\left[-\frac{(\alpha-\alpha_{\rm{mean}})^2}{2\sigma_\alpha^2}\right]}\,.
\end{equation}

\subsection{Results}
In this section, we use the non-detections obtained in this paper in combination with the previously reported detection in the lens system SDSS\,J0946+1006
to derive statistical constraints on the substructure mass function within a region of about 32 kpc$^2$ on average around the Einstein radius, for substructure masses between $4\times10^6M_\odot$ and  $4\times10^9M_\odot$. The results of this analysis are shown in Fig. \ref{fig:mass_function}.

 For a uniform prior probability density on $\alpha$, we find a mean substructure projected mass fraction of $f =~$\ufr ~at the 68 percent CL and a substructure mass function slope $\alpha ~ $\ual ~  at the 95 percent CL.  For a Gaussian prior of mean $1.9$ and standard deviation $0.1$, we infer a fraction of $f =~ $\gfr~ and a slope of $\alpha =~ $\gal ~ at the 68 percent CL (see Table \ref{tbl:clumpy_modeling_results}).  
 
 With a Bayes factor of $2\times\Delta\ln{\cal E} =  0.36$, we find that there is no positive preference for the model with a uniform prior on the slope \citep{Kass95}. This reflects the fact that with only one detection we have poor constraints on the mass function slope, hence only an upper limit is derived. The mass fraction is instead constrained at the lower limit by the single detection and at the upper limit by the non-detections. These results are robust against assumptions on the prior probability density $P(f)$. 
 Thanks to the larger sample of lenses considered in this paper, we can derive tighter constraints on the substructure mass fraction than previously obtained. However, owing to the large number of non-detections, potentially related to the rather low sensitivity of the data analysed here, we still suffer from a large uncertainty on the substructure mass function slope.  \\
\indent Since detailed studies of the luminous satellites of SLACS lens galaxies  have shown that  the substructure found in lenses is representative of that measured around non-lens galaxies \citep{Jackson10, Nirenmberg13a} we are confident that our results can be generalized to the massive elliptical galaxy population. 

\section{Discussion and Conclusions}

\citet{Vegetti10} reported from the analysis of the single detection in the lens system SDSS\,J0946+1006 a substructure mass fraction of $f = 0.0256^{+0.0326}_{-0.0150}$ and $f = 0.0215^{+0.0205}_{-0.0125}$ for a uniform prior and a Gaussian prior on $\alpha$, respectively. This is higher but consistent within the error with the fraction reported in this paper. The mean value of our current analysis is lower due to the large number of non-detections which tightly constrains the fraction upper limit. From the joint analysis of the lens system SDSS\,J0946+1006 and B1938+666, \citet{Vegetti12} derived instead a fraction of $f=0.039^{+0.036}_{-0.024}$  and $f =0.015^{+0.015}_{-0.009, }$ for a uniform prior and a Gaussian prior on $\alpha$, respectively. This results is also larger but consistent with our constraints and this difference is also due to the different number of detections and non-detections considered. Thanks to a larger number of detections, \citet{Vegetti12} were also able to set a constraint on the mass function slope, $\alpha = 1.0^{+0.8}_{-0.1}$. In this paper, we have not included the detection in the lens system B1938+666 in order to keep the lens sample uniform in both lens galaxy mass and redshift. In the future, larger samples of strong gravitational lens galaxies will allow us to constrain the substructure mass function as a function of the host galaxy mass and redshift and to tightly constraint both the mass function slope and normalization.

Our current results imply a fraction which is consistent within the error with the mass fraction derived by \citet{Dalal02} ($0.006 < f < 0.07$) using measurements of flux-ratio anomalies in multiply imaged quasars. It should be kept in mind, however, that the lens sample analysed by \citet{Dalal02} has a mean redshift of 0.6, hence a larger mass fraction should be expected relatively to our z=0.2 sample \citep{Xu13}.

Using rescaled versions of the Aquarius \citep{Springel08} and Phoenix simulations \citep{Gao12}, \citet{Xu13} have explored the effect of mass substructure on the flux ratio of multiply imaged quasars in galaxies with masses, ellipticity and redshift comparable to the observed lens galaxy population. While the effect of substructures on the flux of point sources is different from their effect on the surface brightness distribution of extended sources (the former is a function of the second derivative of the potential, while the latter is a function of its first derivative) we can still use some of the \citet{Xu13} results to compare our findings with the expectations from the CDM model. In particular, \citet{Xu13} have found the number density of subhaloes in the inner regions of the host haloes to be constant and the subhaloes surface mass density to be roughly $4.72\times10^6 h^{-1}M_\odot (h^{-1} \rm{kpc})^{-2}$, for subhalo masses between $5.48 \times10^6M_\odot h^{-1}$ and $5.48 \times10^9M_\odot h^{-1}$, a host halo mass of about 10$^{13}h^{-1}M_\odot$ and a host halo redshift of $z=0.2$. This implies an average substructure mass fraction of $f_{\rm{CDM}}\approx 0.0046$, for substructure masses between $4\times10^6M_\odot$ and  $4\times10^9M_\odot$, and for substructure positions within the same regions considered in this paper. This is consistent with our results at the 68 percent CL. We caution, however, that higher quality data are required in order to extend our measurements to lower substructure masses \citep{Vegetti12, Laga12}, where larger discrepancies between the data and the model are expected. Also, high-resolution hydrodynamical numerical simulations are required for a proper comparison that takes into account for the effect of baryons on the survival of mass substructure within massive host galaxies.

\citet{Vegetti12} have shown that the substructure mass derived under the assumption of a pseudo-Jaffe profile is consisted with the mass derived from the pixelized density corrections. This clearly indicates that the measured substructure mass is a robust quantity that is not significantly affected by the choice of mass profile. Using mock observations of the lens system B1938+666, \citet{Vegetti14} have shown that the assumed profile does not lead to a measurement error on the lensing mass, but that profiles with too low concentrations would not be detectable with the gravitational imaging technique. In practice, this implies that while the detections are not affected by the choice of mass profile, the interpretation of the non-detections could be profile-dependent. This is an important issue and in a future paper we will thoroughly investigate the potential biases related to assumptions on the substructure mass profile in different observational scenarios. \citet{Vegetti10} and \citet{Vegetti12} have shown instead that the galaxy subtraction procedure does not affect the substructure detections and their interpretation.

In conclusion, we have applied the gravitational imaging technique by \citet{Vegetti09a} to a sample of 11 gravitational lens galaxies from the SLACS survey with the aim of constraining the substructure mass function for host lens galaxies at a mean redshift of 0.2
and with a mean velocity dispersion of 270 km\,s$^{-1}$. Our main results can be summarized as follows:

\begin{enumerate}
\item given our criteria for the significant detection of a mass substructure, no new dark mass substructure has been identified;

\item for each lens we have calculated the mass sensitivity function as a function of the substructure position on the lensed image plane within relatively tight regions around the critical curves. We have found that, given the quality of the considered data, the lowest detectable mass typically varies on average between $0.04\times10^{10}M_\odot$ and $0.14\times10^{10}M_\odot$ depending on the lens. In some cases, for specific substructure positions the detection threshold, it can be as low as $0.01\times10^{10}M_\odot$;

\item by statistically combining these non-detections with the detection of a dark-matter-dominated substructure in the lens system SDSS\,J0946+1006 \citep{Vegetti10}, we have inferred a projected mass fraction in substructure which is consistent with observations based on flux-ratio anomalies of multiply imaged quasars \citep{Dalal02} and with the predictions from $N$-body CDM simulations \citep{Xu13} at the 68 percent CL. 
\end{enumerate}

In the near future, thanks to high resolution adaptive optics \citep[e.g. from the Strong lensing at High Angular Resolution Programme, SHARP,][]{Laga12} and to highly structured Einstein rings observed with the {\it HST} in the ultra-violet (e.g. {\it HST} observing program 12898), we expect to extend our sensitivity to substructure masses as low as $10^6 M_\odot$, and therefore test the CDM paradigm in a mass regime where most of the discrepancy between the CDM predicted subhalo mass function and the observed luminosity function is found.

\section*{Acknowledgements}
During part of this work SV acknowledged the support of a Pappalardo Fellowship at the Massachusetts Institute of Technology.
TT acknowledges support from the Packard Foundation in the form of a Packard Research Fellowship. 
Support for programs 10494, 10798, and 11202 was provided by NASA through grants from the Space Telescope Science Institute.
SV is grateful to Paul Schechter and John McKean for insightful discussions, Anna Nierenberg and Sherry Suyu for discussions on the substructure spatial distribution and truncation radius and Dandan Xu for discussions on the comparison between observations and expectations. SV and MWA are grateful to Mustafa Amin for the analytic solutions of the Einstein radius. LVEK, SV and TT thank the Kavli Institute for Theoretical Physics for hosting the 2012 \emph{First Galaxies and Faint Dwarfs} workshop, during which important parts of this work were made. This work is based on observations made with the NASA/ESA {\it Hubble Space Telescope}, obtained from the data archive at the Space Telescope Science Institute. STScI is operated by the Association of Universities for Research in Astronomy, Inc. under NASA contract NAS 5-26555.
\bibliography{ms}

\begin{thebibliography}{}

\bibitem[\protect\citeauthoryear{{Auger}, {Treu}, {Bolton}, {Gavazzi},
  {Koopmans}, {Marshall}, {Bundy} \& {Moustakas}}{{Auger}
  et~al.}{2009}]{Auger09}
{Auger} M.~W.,  {Treu} T.,  {Bolton} A.~S.,  {Gavazzi} R.,  {Koopmans}
  L.~V.~E.,  {Marshall} P.~J.,  {Bundy} K.,    {Moustakas} L.~A.,  2009, \apj,
  705, 1099

\bibitem[\protect\citeauthoryear{{Auger}, {Treu}, {Bolton}, {Gavazzi},
  {Koopmans}, {Marshall}, {Moustakas} \& {Burles}}{{Auger}
  et~al.}{2010}]{Auger10}
{Auger} M.~W.,  {Treu} T.,  {Bolton} A.~S.,  {Gavazzi} R.,  {Koopmans}
  L.~V.~E.,  {Marshall} P.~J.,  {Moustakas} L.~A.,    {Burles} S.,  2010, \apj,
  724, 511

\bibitem[\protect\citeauthoryear{{Barnab{\`e}}, {Czoske}, {Koopmans}, {Treu} \&
  {Bolton}}{{Barnab{\`e}} et~al.}{2011}]{Barnabe11}
{Barnab{\`e}} M.,  {Czoske} O.,  {Koopmans} L.~V.~E.,  {Treu} T.,    {Bolton}
  A.~S.,  2011, \mnras, 415, 2215

\bibitem[\protect\citeauthoryear{{Barnab{\`e}}, {Spiniello}, {Koopmans},
  {Trager}, {Czoske} \& {Treu}}{{Barnab{\`e}} et~al.}{2013}]{Barnabe13}
{Barnab{\`e}} M.,  {Spiniello} C.,  {Koopmans} L.~V.~E.,  {Trager} S.~C.,
  {Czoske} O.,    {Treu} T.,  2013, \mnras, 436, 253

\bibitem[\protect\citeauthoryear{Binney \& Tremaine}{Binney \&
  Tremaine}{1987}]{Binney87}
Binney J.,  Tremaine S.,  1987, Galactic Dynamics.
Princeton University Press

\bibitem[\protect\citeauthoryear{{Bolton}, {Burles}, {Koopmans}, {Treu},
  {Gavazzi}, {Moustakas}, {Wayth} \& {Schlegel}}{{Bolton}
  et~al.}{2008}]{Bolton08a}
{Bolton} A.~S.,  {Burles} S.,  {Koopmans} L.~V.~E.,  {Treu} T.,  {Gavazzi} R.,
  {Moustakas} L.~A.,  {Wayth} R.,    {Schlegel} D.~J.,  2008, \apj, 682, 964

\bibitem[\protect\citeauthoryear{{Bolton}, {Burles}, {Koopmans}, {Treu} \&
  {Moustakas}}{{Bolton} et~al.}{2006}]{Bolton06}
{Bolton} A.~S.,  {Burles} S.,  {Koopmans} L.~V.~E.,  {Treu} T.,    {Moustakas}
  L.~A.,  2006, \apj, 638, 703

\bibitem[\protect\citeauthoryear{{Bolton}, {Treu}, {Koopmans}, {Gavazzi},
  {Moustakas}, {Burles}, {Schlegel} \& {Wayth}}{{Bolton}
  et~al.}{2008}]{Bolton08b}
{Bolton} A.~S.,  {Treu} T.,  {Koopmans} L.~V.~E.,  {Gavazzi} R.,  {Moustakas}
  L.~A.,  {Burles} S.,  {Schlegel} D.~J.,    {Wayth} R.,  2008, \apj, 684, 248

\bibitem[\protect\citeauthoryear{{Brada{\v c}}, {Schneider}, {Steinmetz},
  {Lombardi}, {King} \& {Porcas}}{{Brada{\v c}} et~al.}{2002}]{Bradac02}
{Brada{\v c}} M.,  {Schneider} P.,  {Steinmetz} M.,  {Lombardi} M.,  {King}
  L.~J.,    {Porcas} R.,  2002, \aap, 388, 373

\bibitem[\protect\citeauthoryear{{Cappellari}, {McDermid}, {Alatalo}, {Blitz},
  {Bois}, {Bournaud}, {Bureau}, {Crocker}, {Davies} et~al.,}{{Cappellari}
  et~al.}{2012}]{Cappellari12}
{Cappellari} M.,  {McDermid} R.~M.,  {Alatalo} K.,  {Blitz} L.,  {Bois} M.,
  {Bournaud} F.,  {Bureau} M.,  {Crocker} A.~F.,  {Davies} R.~L.,    et~al.,
  2012, \nat, 484, 485

\bibitem[\protect\citeauthoryear{{Chen}, {Rozo}, {Dalal} \& {Taylor}}{{Chen}
  et~al.}{2007}]{Chen07}
{Chen} J.,  {Rozo} E.,  {Dalal} N.,    {Taylor} J.~E.,  2007, \apj, 659, 52

\bibitem[\protect\citeauthoryear{{Chiba}}{{Chiba}}{2002}]{Chiba02}
{Chiba} M.,  2002, \apj, 565, 17

\bibitem[\protect\citeauthoryear{{Dalal} \& {Kochanek}}{{Dalal} \&
  {Kochanek}}{2002}]{Dalal02}
{Dalal} N.,  {Kochanek} C.~S.,  2002, \apj, 572, 25

\bibitem[\protect\citeauthoryear{Diemand, Kuhlen, Madau, Zemp, Moore, Potter \&
  Stadel}{Diemand et~al.}{2008}]{Diemand08}
Diemand J.,  Kuhlen M.,  Madau P.,  Zemp M.,  Moore B.,  Potter D.,    Stadel
  J.,  2008, Nature, 454, 735

\bibitem[\protect\citeauthoryear{{Eichner}, {Seitz} \& {Bauer}}{{Eichner}
  et~al.}{2012}]{Eichner12}
{Eichner} T.,  {Seitz} S.,    {Bauer} A.,  2012, \mnras, 427, 1918

\bibitem[\protect\citeauthoryear{{Eisenstein}, {Annis}, {Gunn}, {Szalay},
  {Connolly}, {Nichol}, {Bahcall}, {Bernardi} et~al.,}{{Eisenstein}
  et~al.}{2001}]{Eisenstein01}
{Eisenstein} D.~J.,  {Annis} J.,  {Gunn} J.~E.,  {Szalay} A.~S.,  {Connolly}
  A.~J.,  {Nichol} R.~C.,  {Bahcall} N.~A.,  {Bernardi} M.,    et~al., 2001,
  \aj, 122, 2267

\bibitem[\protect\citeauthoryear{{Fadely} \& {Keeton}}{{Fadely} \&
  {Keeton}}{2012}]{Fadely12}
{Fadely} R.,  {Keeton} C.~R.,  2012, \mnras, 419, 936

\bibitem[\protect\citeauthoryear{{Feroz} \& {Hobson}}{{Feroz} \&
  {Hobson}}{2008}]{Feroz08}
{Feroz} F.,  {Hobson} M.~P.,  2008, \mnras, 384, 449

\bibitem[\protect\citeauthoryear{{Gao}, {Navarro}, {Frenk}, {Jenkins},
  {Springel} \& {White}}{{Gao} et~al.}{2012}]{Gao12}
{Gao} L.,  {Navarro} J.~F.,  {Frenk} C.~S.,  {Jenkins} A.,  {Springel} V.,
  {White} S.~D.~M.,  2012, \mnras, 425, 2169

\bibitem[\protect\citeauthoryear{{Jackson}, {Bryan}, {Mao} \& {Li}}{{Jackson}
  et~al.}{2010}]{Jackson10}
{Jackson} N.,  {Bryan} S.~E.,  {Mao} S.,    {Li} C.,  2010, \mnras, 403, 826

\bibitem[\protect\citeauthoryear{{Jeffreys}}{{Jeffreys}}{1946}]{Jeffreys46}
{Jeffreys} H.,  1946, Proc. R. Soc. Lond. A, 186, 453

\bibitem[\protect\citeauthoryear{{Kass} \& {Raftery}}{{Kass} \&
  {Raftery}}{1995}]{Kass95}
{Kass} R.,  {Raftery} A.,  1995, Journal of the American Statistical
  Association, 90, 773

\bibitem[\protect\citeauthoryear{{Kauffmann}, {White} \&
  {Guiderdoni}}{{Kauffmann} et~al.}{1993}]{Kauffmann93}
{Kauffmann} G.,  {White} S.~D.~M.,    {Guiderdoni} B.,  1993, \mnras, 264, 201

\bibitem[\protect\citeauthoryear{{Keeton}, {Gaudi} \& {Petters}}{{Keeton}
  et~al.}{2003}]{Keeton03}
{Keeton} C.~R.,  {Gaudi} B.~S.,    {Petters} A.~O.,  2003, \apj, 598, 138

\bibitem[\protect\citeauthoryear{{Keeton} \& {Moustakas}}{{Keeton} \&
  {Moustakas}}{2009}]{Keeton09}
{Keeton} C.~R.,  {Moustakas} L.~A.,  2009, \apj, 699, 1720

\bibitem[\protect\citeauthoryear{{Klypin}, {Kravtsov}, {Valenzuela} \&
  {Prada}}{{Klypin} et~al.}{1999}]{Klypin99}
{Klypin} A.,  {Kravtsov} A.~V.,  {Valenzuela} O.,    {Prada} F.,  1999, \apj,
  522, 82

\bibitem[\protect\citeauthoryear{{Koester}, {McKay}, {Annis}, {Wechsler},
  {Evrard}, {Bleem}, {Becker}, {Johnston},  et~al.,}{{Koester}
  et~al.}{2007}]{Koester07}
{Koester} B.~P.,  {McKay} T.~A.,  {Annis} J.,  {Wechsler} R.~H.,  {Evrard} A.,
  {Bleem} L.,  {Becker} M.,  {Johnston} D.,     et~al., 2007, \apj, 660, 239

\bibitem[\protect\citeauthoryear{{Koopmans}}{{Koopmans}}{2005}]{Koopmans05}
{Koopmans} L.~V.~E.,  2005, \mnras, 363, 1136

\bibitem[\protect\citeauthoryear{{Koopmans}, {Bolton}, {Treu}, {Czoske},
  {Auger}, {Barnab{\`e}}, {Vegetti}, {Gavazzi}, {Moustakas} \&
  {Burles}}{{Koopmans} et~al.}{2009}]{Koopmans09}
{Koopmans} L.~V.~E.,  {Bolton} A.,  {Treu} T.,  {Czoske} O.,  {Auger} M.~W.,
  {Barnab{\`e}} M.,  {Vegetti} S.,  {Gavazzi} R.,  {Moustakas} L.~A.,
  {Burles} S.,  2009, \apjl, 703, L51

\bibitem[\protect\citeauthoryear{{Kormann}, {Schneider} \&
  {Bartelmann}}{{Kormann} et~al.}{1994}]{Kormann94}
{Kormann} R.,  {Schneider} P.,    {Bartelmann} M.,  1994, \aap, 284, 285

\bibitem[\protect\citeauthoryear{{Kravtsov}}{{Kravtsov}}{2010}]{Kravtsov10}
{Kravtsov} A.,  2010, Advances in Astronomy, 2010, 8

\bibitem[\protect\citeauthoryear{{Lagattuta}, {Vegetti}, {Fassnacht}, {Auger},
  {Koopmans} \& {McKean}}{{Lagattuta} et~al.}{2012}]{Laga12}
{Lagattuta} D.~J.,  {Vegetti} S.,  {Fassnacht} C.~D.,  {Auger} M.~W.,
  {Koopmans} L.~V.~E.,    {McKean} J.~P.,  2012, \mnras, 424, 2800

\bibitem[\protect\citeauthoryear{{Mao} \& {Schneider}}{{Mao} \&
  {Schneider}}{1998}]{Mao98}
{Mao} S.,  {Schneider} P.,  1998, \mnras, 295, 587

\bibitem[\protect\citeauthoryear{{McKean}, {Koopmans}, {Flack}, {Fassnacht},
  {Thompson}, {Matthews}, {Blandford}, {Readhead} \& {Soifer}}{{McKean}
  et~al.}{2007}]{McKean07}
{McKean} J.~P.,  {Koopmans} L.~V.~E.,  {Flack} C.~E.,  {Fassnacht} C.~D.,
  {Thompson} D.,  {Matthews} K.,  {Blandford} R.~D.,  {Readhead} A.~C.~S.,
  {Soifer} B.~T.,  2007, \mnras, 378, 109

\bibitem[\protect\citeauthoryear{{Metcalf} \& {Amara}}{{Metcalf} \&
  {Amara}}{2012}]{Metcalf12}
{Metcalf} R.~B.,  {Amara} A.,  2012, \mnras, 419, 3414

\bibitem[\protect\citeauthoryear{{Metcalf} \& {Madau}}{{Metcalf} \&
  {Madau}}{2001}]{Metcalf01}
{Metcalf} R.~B.,  {Madau} P.,  2001, \apj, 563, 9

\bibitem[\protect\citeauthoryear{{Metcalf} \& {Zhao}}{{Metcalf} \&
  {Zhao}}{2002}]{Metcalf02}
{Metcalf} R.~B.,  {Zhao} H.,  2002, \apjl, 567, L5

\bibitem[\protect\citeauthoryear{{More}, {McKean}, {More}, {Porcas}, {Koopmans}
  \& {Garrett}}{{More} et~al.}{2009}]{More09}
{More} A.,  {McKean} J.~P.,  {More} S.,  {Porcas} R.~W.,  {Koopmans} L.~V.~E.,
    {Garrett} M.~A.,  2009, \mnras, 394, 174

\bibitem[\protect\citeauthoryear{{Nierenberg}, {Auger}, {Treu}, {Marshall} \&
  {Fassnacht}}{{Nierenberg} et~al.}{2011}]{Nierenberg11}
{Nierenberg} A.~M.,  {Auger} M.~W.,  {Treu} T.,  {Marshall} P.~J.,
  {Fassnacht} C.~D.,  2011, \apj, 731, 44

\bibitem[\protect\citeauthoryear{{Nierenberg}, {Auger}, {Treu}, {Marshall},
  {Fassnacht} \& {Busha}}{{Nierenberg} et~al.}{2012}]{Nierenberg12}
{Nierenberg} A.~M.,  {Auger} M.~W.,  {Treu} T.,  {Marshall} P.~J.,  {Fassnacht}
  C.~D.,    {Busha} M.~T.,  2012, \apj, 752, 99

\bibitem[\protect\citeauthoryear{{Nierenberg}, {Oldenburg} \&
  {Treu}}{{Nierenberg} et~al.}{2013}]{Nirenmberg13a}
{Nierenberg} A.~M.,  {Oldenburg} D.,    {Treu} T.,  2013, \mnras, 436, 2120

\bibitem[\protect\citeauthoryear{{Nierenberg}, {Treu}, {Menci}, {Lu} \&
  {Wang}}{{Nierenberg} et~al.}{2013}]{Nierenberg13}
{Nierenberg} A.~M.,  {Treu} T.,  {Menci} N.,  {Lu} Y.,    {Wang} W.,  2013,
  \apj, 772, 146

\bibitem[\protect\citeauthoryear{{Nierenberg}, {Treu}, {Wright}, {Fassnacht} \&
  {Auger}}{{Nierenberg} et~al.}{2014}]{Nierenberg14}
{Nierenberg} A.~M.,  {Treu} T.,  {Wright} S.~A.,  {Fassnacht} C.~D.,    {Auger}
  M.~W.,  2014, ArXiv e-prints

\bibitem[\protect\citeauthoryear{{Rau}, {Vegetti} \& {White}}{{Rau}
  et~al.}{2013}]{Rau13}
{Rau} S.,  {Vegetti} S.,    {White} S.~D.~M.,  2013, \mnras, 430, 2232

\bibitem[\protect\citeauthoryear{{Schechter} \& {Moore}}{{Schechter} \&
  {Moore}}{1993}]{Schechter93}
{Schechter} P.~L.,  {Moore} C.~B.,  1993, \aj, 105, 1

\bibitem[\protect\citeauthoryear{Skilling}{Skilling}{2004}]{Skilling04}
Skilling J.,  2004, Nested Sampling for General Bayesian Computation.
in Fisher, R. and Preuss, R. and Toussaint, U. V., eds, AIP Conf. Series Vol.
  735. Am. Inst. Phys. Rev. Lett. 84, 3760

\bibitem[\protect\citeauthoryear{{Sonnenfeld}, {Treu}, {Gavazzi}, {Suyu},
  {Marshall}, {Auger} \& {Nipoti}}{{Sonnenfeld} et~al.}{2013}]{Sonnenfeld13}
{Sonnenfeld} A.,  {Treu} T.,  {Gavazzi} R.,  {Suyu} S.~H.,  {Marshall} P.~J.,
  {Auger} M.~W.,    {Nipoti} C.,  2013, \apj, 777, 98

\bibitem[\protect\citeauthoryear{{Spiniello}, {Koopmans}, {Trager}, {Czoske} \&
  {Treu}}{{Spiniello} et~al.}{2011}]{Spiniello11}
{Spiniello} C.,  {Koopmans} L.~V.~E.,  {Trager} S.~C.,  {Czoske} O.,    {Treu}
  T.,  2011, \mnras, 417, 3000

\bibitem[\protect\citeauthoryear{{Springel}, {Wang}, {Vogelsberger}, {Ludlow},
  {Jenkins}, {Helmi}, {Navarro}, {Frenk} \& {White}}{{Springel}
  et~al.}{2008}]{Springel08}
{Springel} V.,  {Wang} J.,  {Vogelsberger} M.,  {Ludlow} A.,  {Jenkins} A.,
  {Helmi} A.,  {Navarro} J.~F.,  {Frenk} C.~S.,    {White} S.~D.~M.,  2008,
  \mnras, 391, 1685

\bibitem[\protect\citeauthoryear{{Strauss}, {Weinberg}, {Lupton}, {Narayanan},
  {Annis}, {Bernardi}, {Blanton}, {Burles} et~al.,}{{Strauss}
  et~al.}{2002}]{Strauss02}
{Strauss} M.~A.,  {Weinberg} D.~H.,  {Lupton} R.~H.,  {Narayanan} V.~K.,
  {Annis} J.,  {Bernardi} M.,  {Blanton} M.,  {Burles} S.,    et~al., 2002,
  \aj, 124, 1810

\bibitem[\protect\citeauthoryear{{Suyu} \& {Halkola}}{{Suyu} \&
  {Halkola}}{2010}]{Suyu10}
{Suyu} S.~H.,  {Halkola} A.,  2010, \aap, 524, A94

\bibitem[\protect\citeauthoryear{{Suyu}, {Marshall}, {Hobson} \&
  {Blandford}}{{Suyu} et~al.}{2006}]{Suyu06}
{Suyu} S.~H.,  {Marshall} P.~J.,  {Hobson} M.~P.,    {Blandford} R.~D.,  2006,
  \mnras, 371, 983

\bibitem[\protect\citeauthoryear{{Treu}, {Auger}, {Koopmans}, {Gavazzi},
  {Marshall} \& {Bolton}}{{Treu} et~al.}{2010}]{Treu10}
{Treu} T.,  {Auger} M.~W.,  {Koopmans} L.~V.~E.,  {Gavazzi} R.,  {Marshall}
  P.~J.,    {Bolton} A.~S.,  2010, \apj, 709, 1195

\bibitem[\protect\citeauthoryear{{Vegetti} \& {Koopmans}}{{Vegetti} \&
  {Koopmans}}{2009a}]{Vegetti09a}
{Vegetti} S.,  {Koopmans} L.~V.~E.,  2009a, \mnras, 392, 945

\bibitem[\protect\citeauthoryear{{Vegetti} \& {Koopmans}}{{Vegetti} \&
  {Koopmans}}{2009b}]{Vegetti09b}
{Vegetti} S.,  {Koopmans} L.~V.~E.,  2009b, \mnras, 400, 1583

\bibitem[\protect\citeauthoryear{{Vegetti}, {Koopmans}, {Bolton}, {Treu} \&
  {Gavazzi}}{{Vegetti} et~al.}{2010}]{Vegetti10}
{Vegetti} S.,  {Koopmans} L.~V.~E.,  {Bolton} A.,  {Treu} T.,    {Gavazzi} R.,
  2010, \mnras, 408, 1969

\bibitem[\protect\citeauthoryear{{Vegetti}, {Lagattuta}, {McKean}, {Auger},
  {Fassnacht} \& {Koopmans}}{{Vegetti} et~al.}{2012}]{Vegetti12}
{Vegetti} S.,  {Lagattuta} D.~J.,  {McKean} J.~P.,  {Auger} M.~W.,  {Fassnacht}
  C.~D.,    {Koopmans} L.~V.~E.,  2012, \nat, 481, 341

\bibitem[\protect\citeauthoryear{{Vegetti} \& {Vogelsberger}}{{Vegetti} \&
  {Vogelsberger}}{2014}]{Vegetti14}
{Vegetti} S.,  {Vogelsberger} M.,  2014, \mnras, accepted

\bibitem[\protect\citeauthoryear{{Wang}, {Sales}, {Henriques} \&
  {White}}{{Wang} et~al.}{2014}]{Wang14}
{Wang} W.,  {Sales} L.~V.,  {Henriques} B.~M.~B.,    {White} S.~D.~M.,  2014,
  ArXiv e-prints

\bibitem[\protect\citeauthoryear{{Wayth} \& {Webster}}{{Wayth} \&
  {Webster}}{2006}]{Wayth06}
{Wayth} R.~B.,  {Webster} R.~L.,  2006, \mnras, 372, 1187

\bibitem[\protect\citeauthoryear{{Xu}, {Mao}, {Cooper}, {Gao}, {Frenk},
  {Angulo} \& {Helly}}{{Xu} et~al.}{2012}]{Xu12}
{Xu} D.~D.,  {Mao} S.,  {Cooper} A.~P.,  {Gao} L.,  {Frenk} C.~S.,  {Angulo}
  R.~E.,    {Helly} J.,  2012, \mnras, 421, 2553

\bibitem[\protect\citeauthoryear{{Xu}, {Mao}, {Wang}, {Springel}, {Gao},
  {White}, {Frenk}, {Jenkins}, {Li} \& {Navarro}}{{Xu} et~al.}{2009}]{Xu09}
{Xu} D.~D.,  {Mao} S.,  {Wang} J.,  {Springel} V.,  {Gao} L.,  {White}
  S.~D.~M.,  {Frenk} C.~S.,  {Jenkins} A.,  {Li} G.,    {Navarro} J.~F.,  2009,
  \mnras, 398, 1235

\bibitem[\protect\citeauthoryear{{Xu}, {Sluse}, {Gao}, {Wang}, {Frenk}, {Mao}
  \& {Schneider}}{{Xu} et~al.}{2013}]{Xu13}
{Xu} D.~D.,  {Sluse} D.,  {Gao} L.,  {Wang} J.,  {Frenk} C.,  {Mao} S.,
  {Schneider} P.,  2013, ArXiv e-prints

\bibitem[\protect\citeauthoryear{{York}, {Adelman}, {Anderson} Jr., {Anderson},
  {Annis}, {Bahcall}, {Bakken}, {Barkhouser},  et~al.,}{{York}
  et~al.}{2000}]{York00}
{York} D.~G.,  {Adelman} J.,  {Anderson} Jr. J.~E.,  {Anderson} S.~F.,  {Annis}
  J.,  {Bahcall} N.~A.,  {Bakken} J.~A.,  {Barkhouser} R.,     et~al., 2000,
  \aj, 120, 1579

\end{thebibliography}


\begin{figure*}
\begin{center} 
\subfigure{
\includegraphics[width=14 cm]{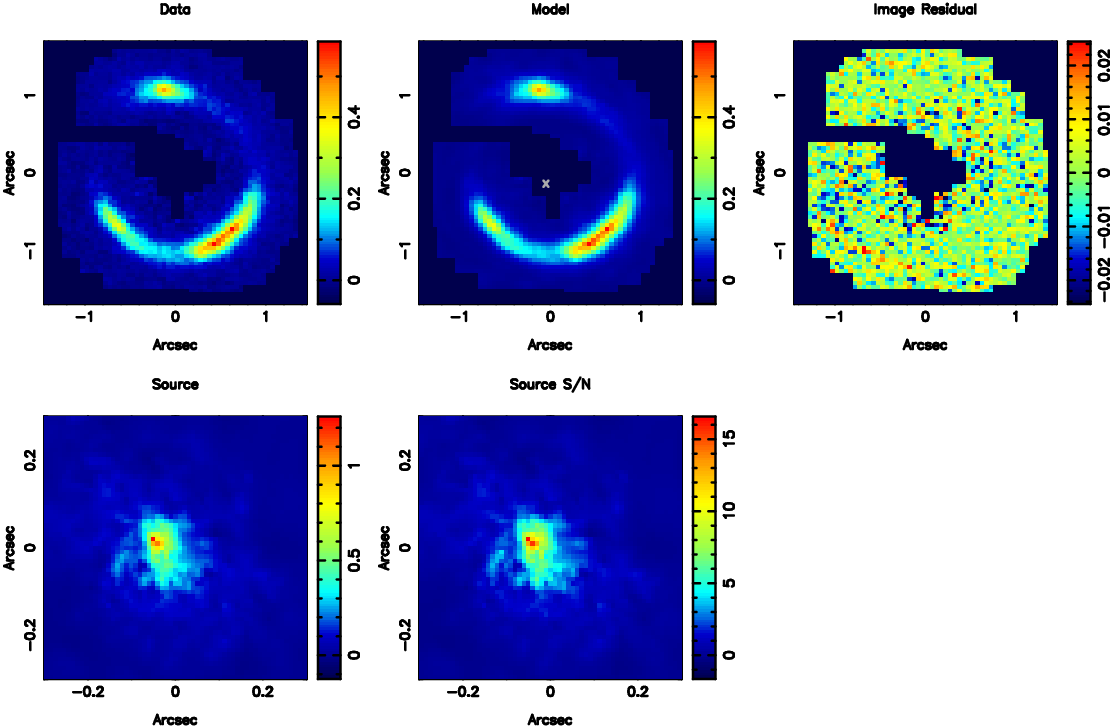}}
\subfigure{
\includegraphics[width=14 cm]{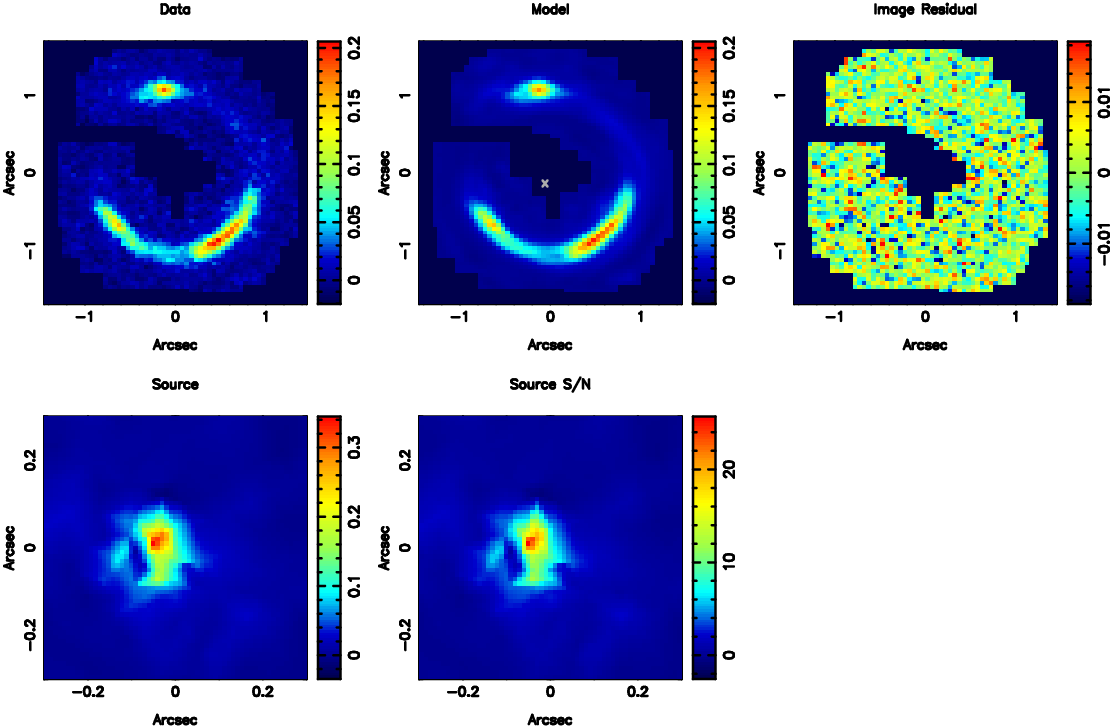}}
\caption{Same as figure \ref{fig:J0252_smooth} for J0737+3216 as observed in the HST F814W filter (top) and in the HST F555W filter (bottom).}
\end{center}     
\end{figure*}	

\begin{figure*}
\begin{center} 
\subfigure{
\includegraphics[width=14 cm]{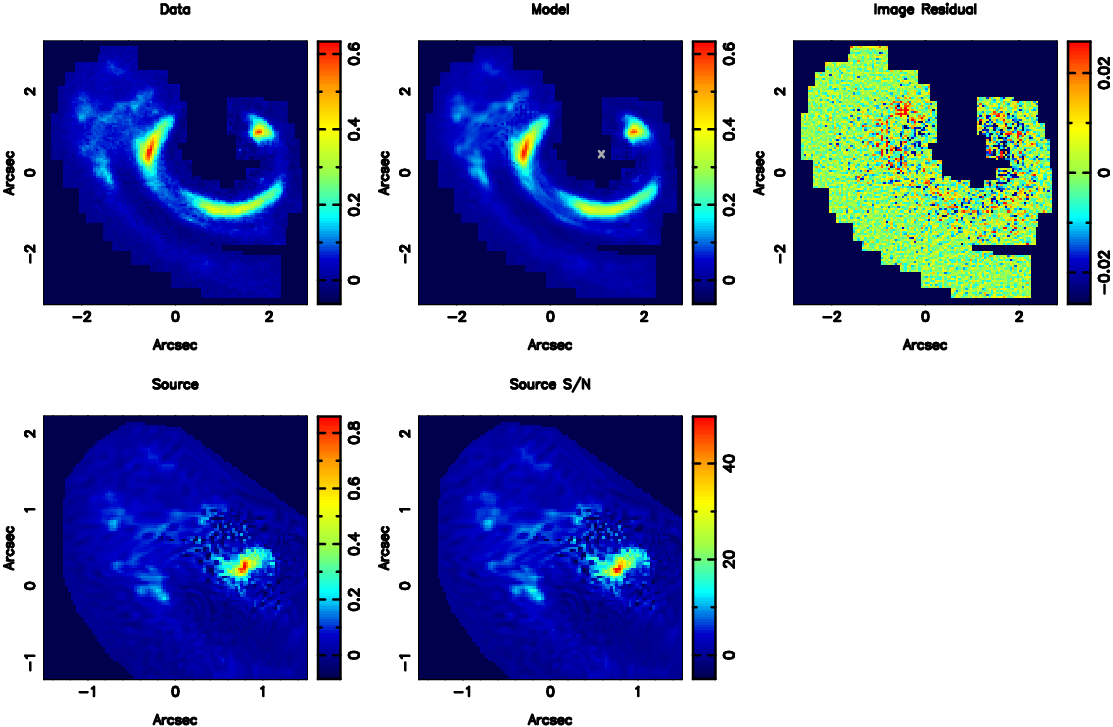}}
\subfigure{
\includegraphics[width=14 cm]{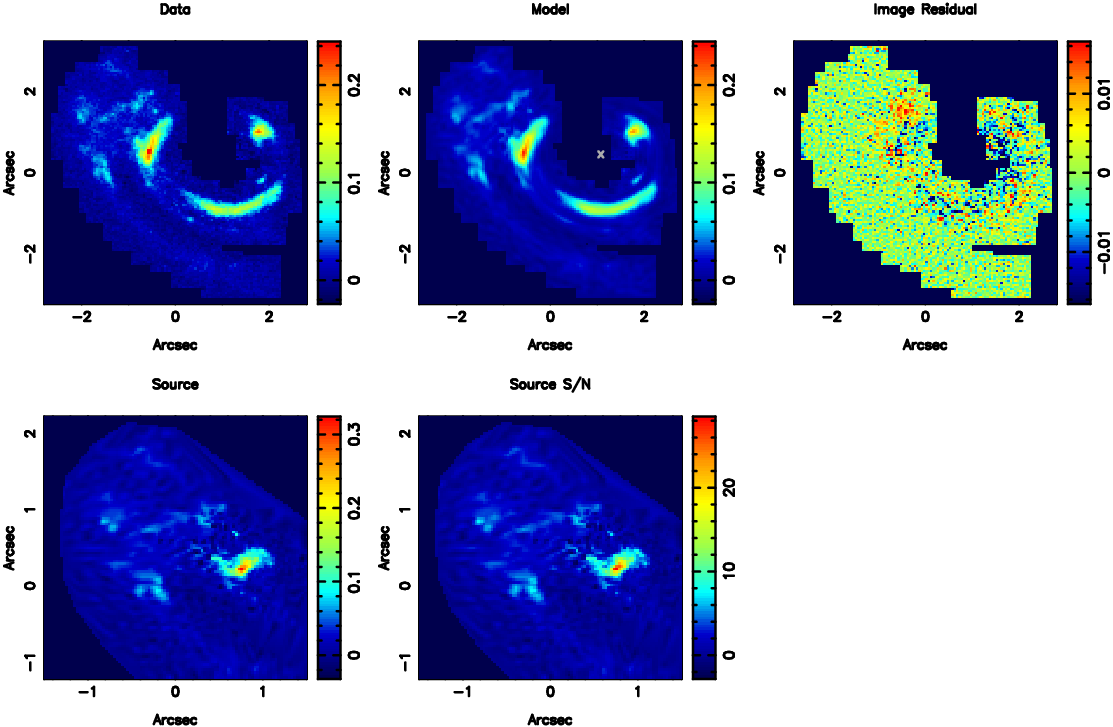}}
\caption{Same as figure \ref{fig:J0252_smooth} for J0956+5100 as observed in the HST F814W filter (top) and in the HST F555W filter (bottom).}
\end{center}     
\end{figure*}

\begin{figure*}
\begin{center} 
\includegraphics[width=14 cm]{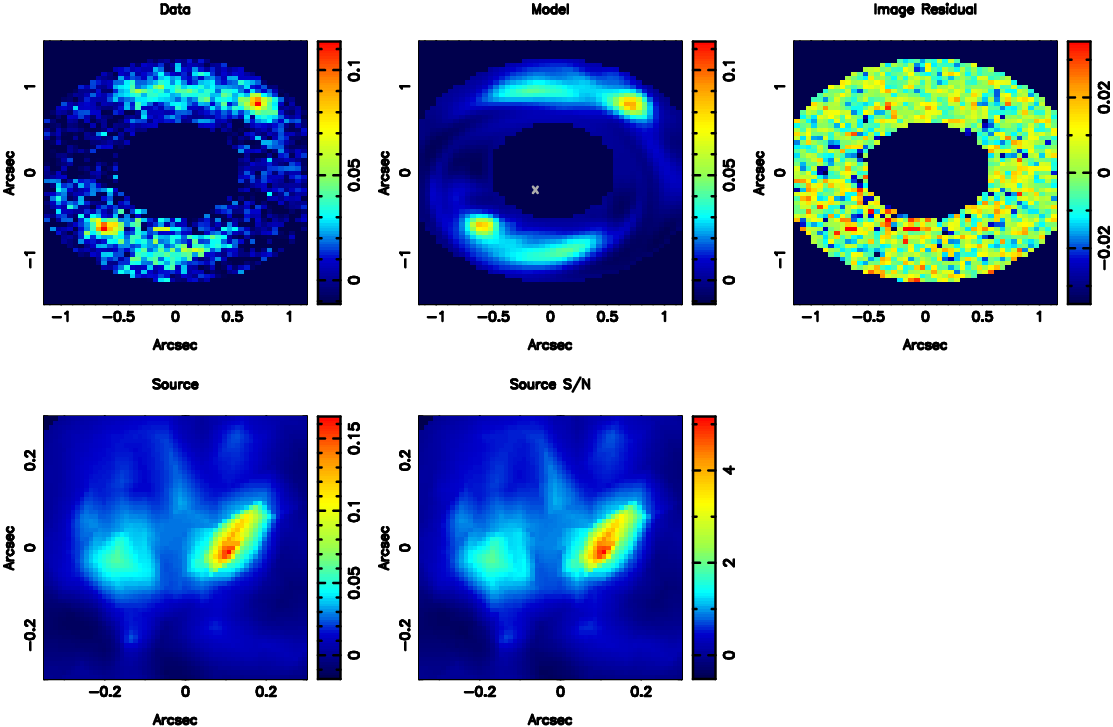}
\caption{Same as figure \ref{fig:J0252_smooth} for J0959+4416.}
\end{center}     
\end{figure*}

\begin{figure*}
\begin{center} 
\includegraphics[width=14 cm]{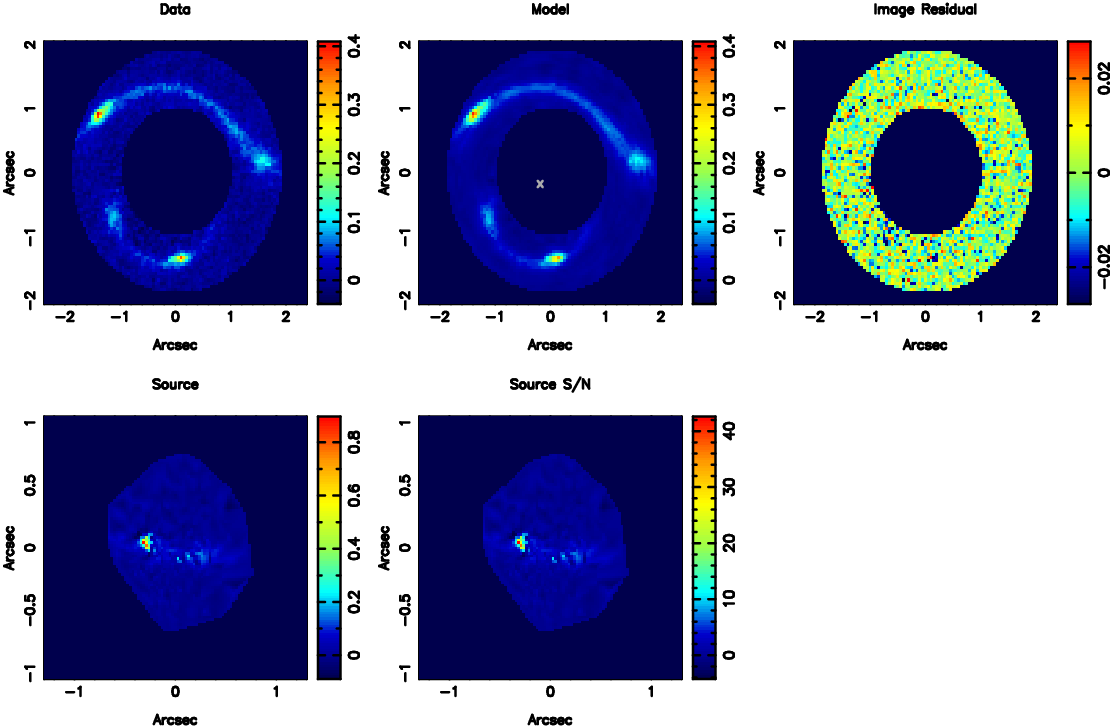}
\caption{Same as figure \ref{fig:J0252_smooth} for J1023+4230.}
\end{center}     
\end{figure*}

\begin{figure*}
\begin{center} 
\subfigure{
\includegraphics[width=14 cm]{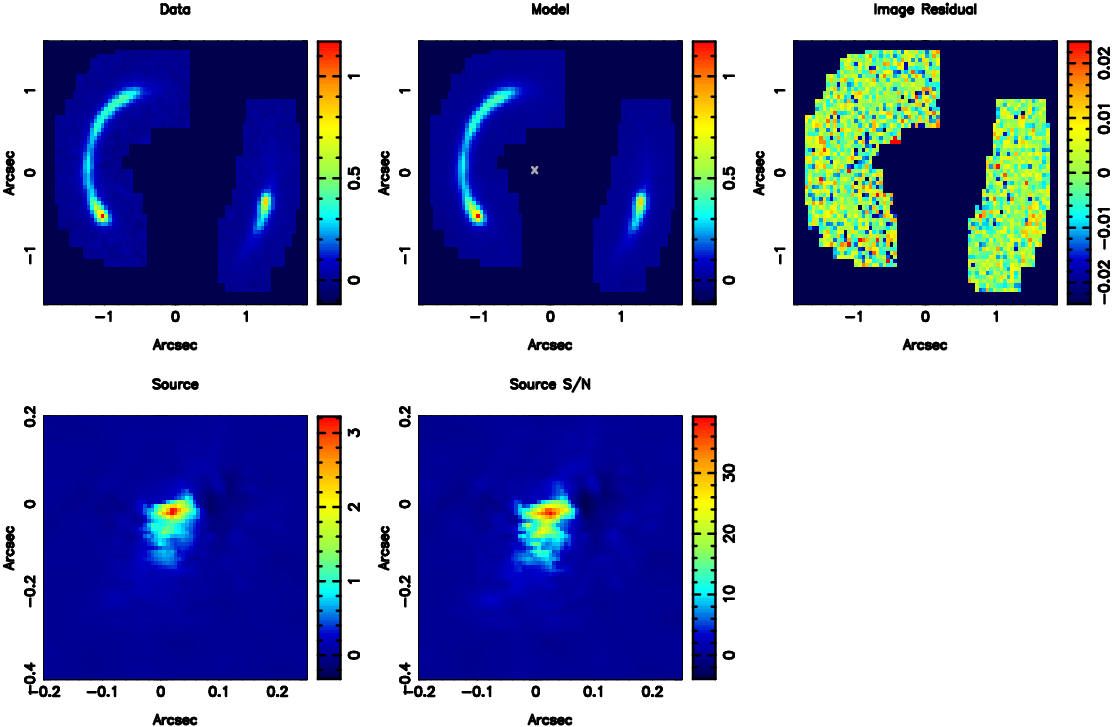}}
\subfigure{
\includegraphics[width=14 cm]{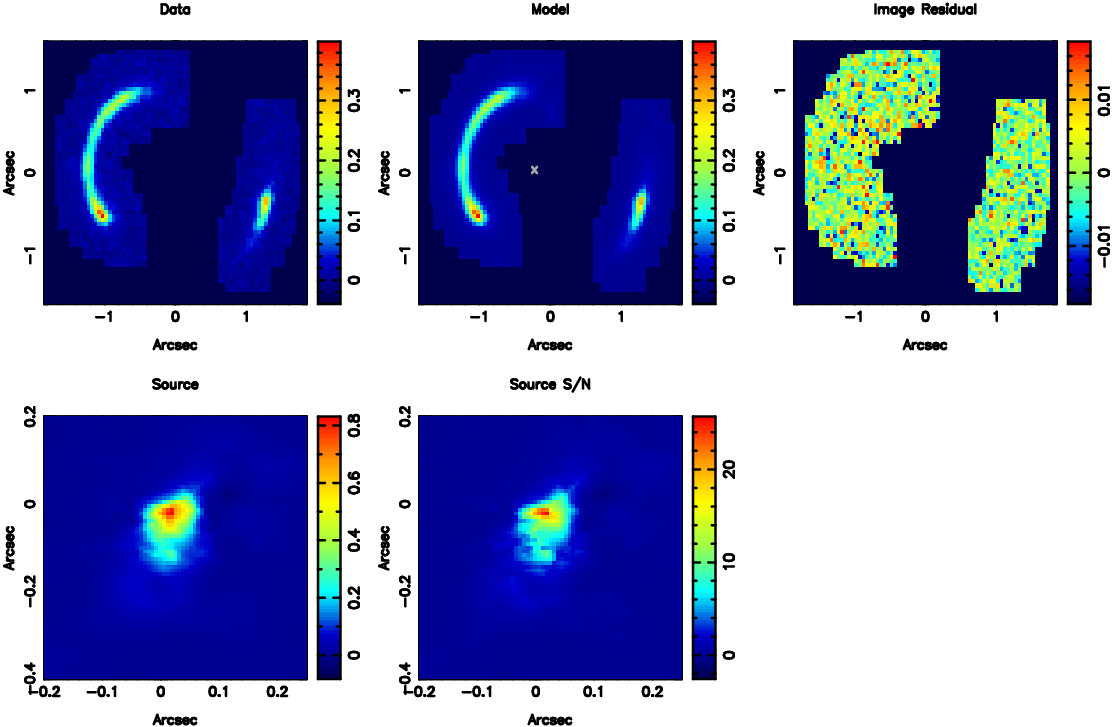}}
\caption{Same as figure \ref{fig:J0252_smooth} for J1205+4910 as observed in the HST F814W filter (top) and in the HST F555W filter (bottom).}
\end{center}     
\end{figure*}

\begin{figure*}
\begin{center} 
\includegraphics[width=14 cm]{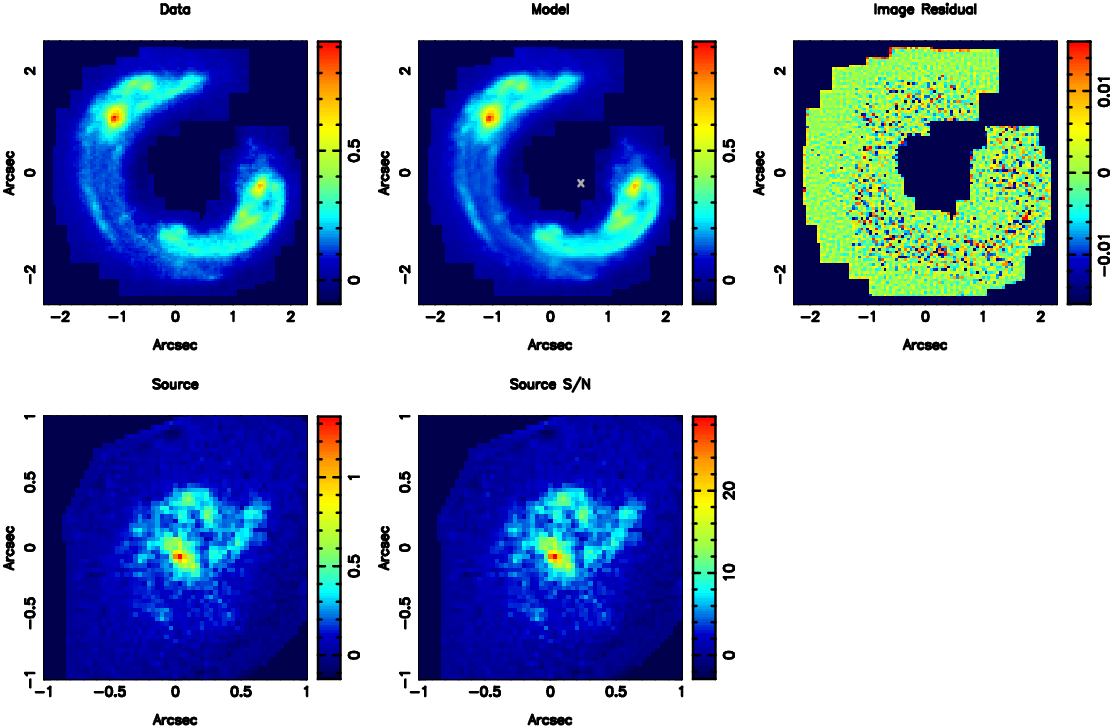}
\caption{Same as figure \ref{fig:J0252_smooth} for J1430+4105.}
\end{center}     
\end{figure*}

\begin{figure*}
\begin{center} 
\subfigure{
\includegraphics[width=14 cm]{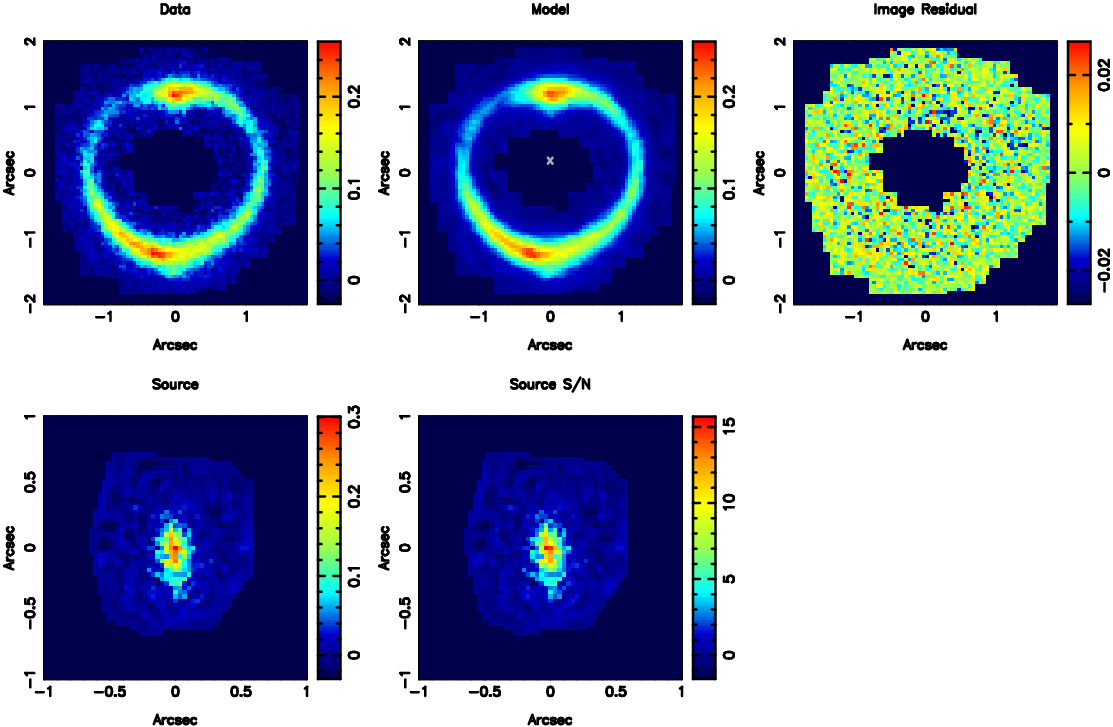}}
\subfigure{
\includegraphics[width=14 cm]{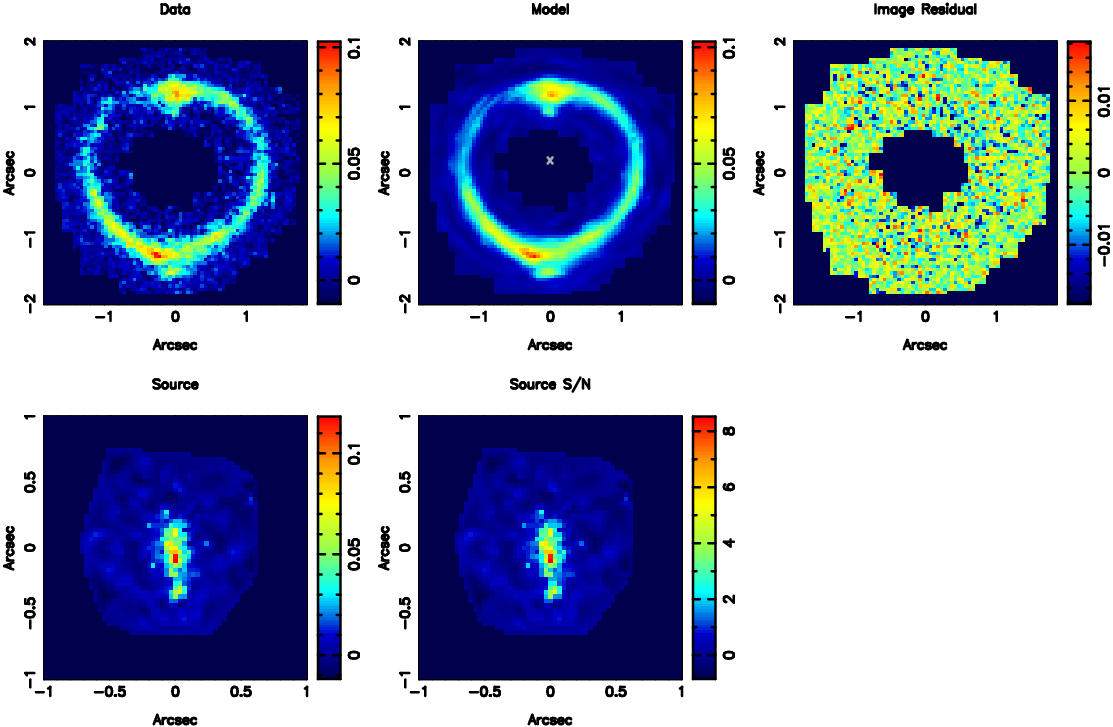}}
\caption{Same as figure \ref{fig:J0252_smooth} for J1627-0053 as observed in the HST F814W filter (top) and in the HST F555W filter (bottom).}
\end{center}     
\end{figure*}

\begin{figure*}
\begin{center} 
\subfigure{
\includegraphics[width=14 cm]{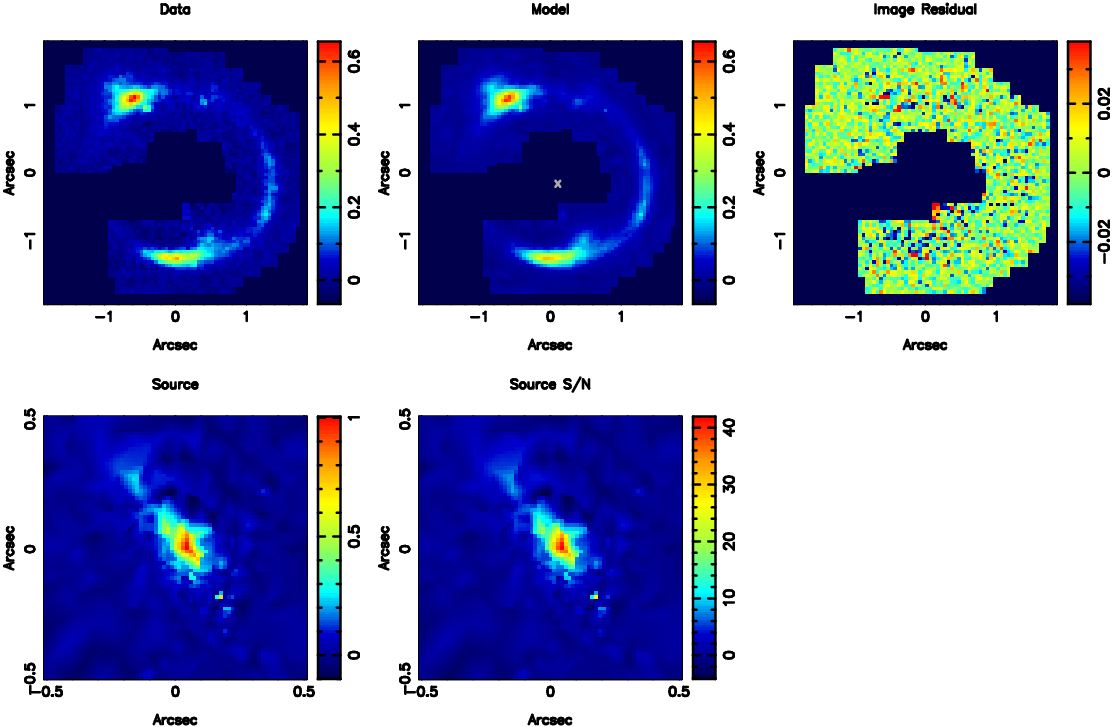}}
\subfigure{
\includegraphics[width=14 cm]{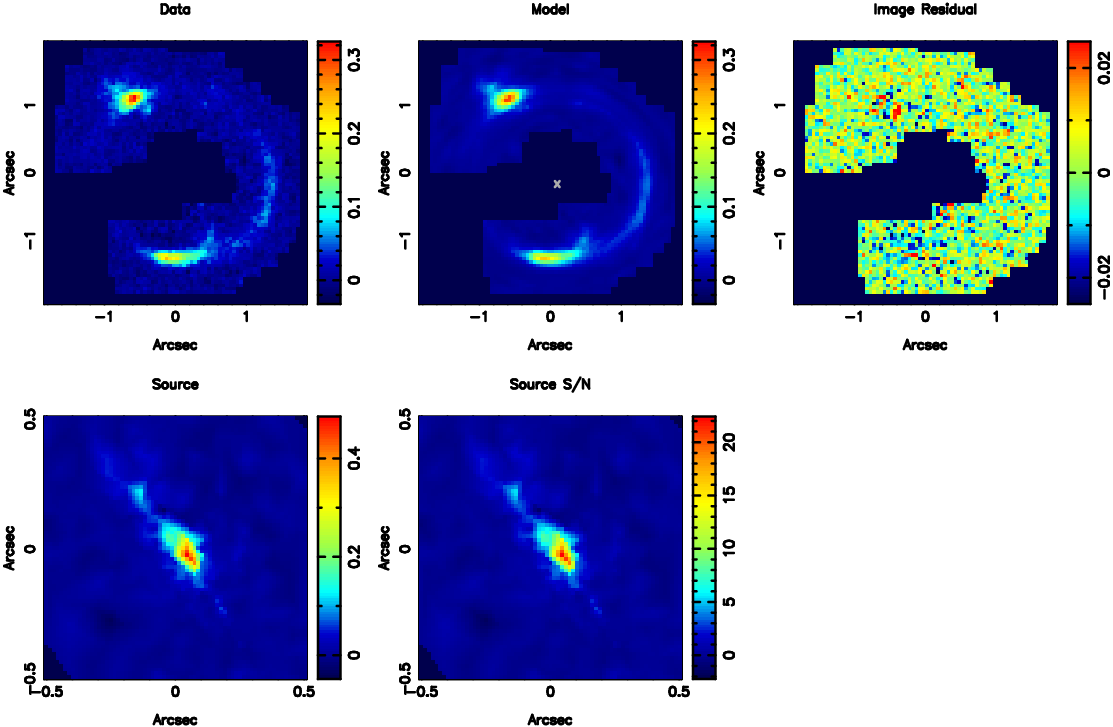}}
\caption{Same as figure \ref{fig:J0252_smooth} for J2238-0754 as observed in the HST F814W filter (top) and in the HST F555W filter (bottom).}
\end{center}     
\end{figure*}

\end{document}